\documentclass[aps,prx,reprint,superscriptaddress,amsmath,amssymb]{revtex4-2}

\usepackage{graphicx}
\usepackage{dcolumn}
\usepackage{bm}
\usepackage{amsmath}
\usepackage{amssymb}
\usepackage{latexsym}
\usepackage{epsfig}
\usepackage{amsbsy}
\usepackage{array}
\usepackage{amssymb}
\usepackage{setspace}
\usepackage{mathtools}
\usepackage{amsthm}
\usepackage{physics}

\usepackage{amsmath,amssymb,amsthm,yhmath,amsfonts,mathrsfs,enumitem,mathtools,dsfont}
\usepackage[utf8]{inputenc}
\usepackage{tikz}
\usetikzlibrary{shapes,arrows,positioning}

\usepackage{graphicx}
\usepackage{tikz}
\usetikzlibrary{automata,positioning}

\usepackage{amsthm}

\usepackage{apptools}
\AtAppendix{\counterwithin{lemma}{section}}
\AtAppendix{\counterwithin{theorem}{section}}

\usepackage{algpseudocode}

\usepackage[justification=centerlast]{caption}
\usepackage[labelformat=simple]{subcaption}

    \usepackage[justification=centerlast]{caption}

\usepackage{xcolor}

\definecolor{orange}{HTML}{B67352}

\begin{document}

\title{Path Percolation in Quantum Communication Networks}

\author{Xiangyi Meng}%
\thanks{These authors contributed equally.}
\affiliation{Department of Physics and Astronomy, Northwestern University, Evanston, Illinois 60208, USA}%

\author{Bingjie Hao}%
\thanks{These authors contributed equally.}
\affiliation{Department of Physics and Astronomy, Northwestern University, Evanston, Illinois 60208, USA}%

\author{Bal\'azs R\'ath}%
\affiliation{Department of Stochastics, Institute of Mathematics, Budapest University of Technology and Economics, H-1111 Budapest, Hungary} 
\affiliation{HUN-REN-BME Stochastics Research Group, H-1111 Budapest, Hungary}
\affiliation{Alfr\'ed R\'enyi Institute of Mathematics, 1053 Budapest, Hungary}

\author{István~A.~Kovács}%
\email{istvan.kovacs@northwestern.edu}
\affiliation{Department of Physics and Astronomy, Northwestern University, Evanston, Illinois 60208, USA}%
\affiliation{Northwestern Institute on Complex Systems, Northwestern University, Evanston, Illinois 60208, USA}
\affiliation{Department of Engineering Sciences and Applied Mathematics, Northwestern University, Evanston, Illinois 60208, USA}

\date{\today}

\begin{abstract}
In a quantum communication network, links represent entanglement between qubits located at different nodes. Even if two nodes are not directly linked by shared entanglement, communication channels can be established between them via quantum routing protocols. However, in contrast to classical communication networks, each communication event removes all participating links along the communication path, disrupting the quantum network. Here, we propose a simple model, where randomly selected pairs of nodes communicate through shortest paths. Each time such a path is used, all participating links are eliminated, leading to a correlated percolation process that we call ``path percolation.'' We study path percolation both numerically and analytically and present the phase diagram of the steady states as a function of the rate at which new links are being added to the quantum communication network. 
As a key result, the steady state is found to be independent from the initial network topology when new link are added randomly between disconnected components.
We close by discussing extensions of path percolation and their potential applications. 
\end{abstract}

\maketitle

The prospect of a quantum Internet heralds a significant shift from classical to quantum communication~\cite{q-internet_k08,*q-internet_weh18}. The latter facilitates applications such as quantum cryptography~\cite{q-crypto_grtz02}, super-dense coding~\cite{q-superdense_bw92}, and cloud quantum computing~\cite{Dumitrescu2018}, promising unprecedented security and efficiency beyond the scope of its classical counterpart. Yet, in contrast to classical communication, quantum communication demands both classical channels and \emph{entanglement}~\cite{q-commun_gt07}---a critical quantum resource~\cite{q-resour_cg19} that must be established, stored (if quantum memories are available \cite{meng2024quantum}), and consumed during quantum communication~\cite{q-resour_kl21}. 
Here, we consider a large-scale quantum communication network of $N$ users (nodes). In such a network, each link denotes a perfectly entangled pair (ebit) established between two nodes. 
As $N$ increases, it will quickly become unfeasible to add a direct link to every pair of nodes~\cite{netw-sparse_gdd24}.
In the case when two nodes do not share a direct link but rather a path comprising \emph{several} links, these links are all consumed by entanglement swapping~\cite{entangle-swap_zzhe93,*entangle-swap_rmkvschb08} when the two nodes communicate one qubit of information. As a result, the consumption of ebits occurs not per link, but \emph{per path}, suggesting that communication of $1$~qubit 
incurs a significant resource cost of 
$l$~ebits, where $l$ is the length of the (potentially shortest) path between two uniformly randomly chosen nodes, both of which must be within the same connected component. 
The inherent removal of paths departs from random link failures~\cite{netw-percolation_ceah00} or edge-cut attacks~\cite{netw-attack_hkyh02}, resulting in a \emph{dynamic} fragmentation process that we term as ``path percolation.''
In this work, we explore the problem of path percolation, focusing on addressing the following related questions:
(i) how many pairs of nodes can communicate before the network falls apart? (ii) if we allow link replenishment, i.e.,~$\alpha \geq 1$ links are generated in the network per time step, then what are the possible steady states of the system?

\begin{figure}[t!]
    \centering
    \begin{minipage}[b]{121pt}
		\centering
		{\subcaption{\label{fig_demo_remove}}\includegraphics[width=121pt]{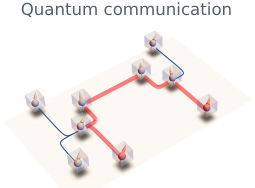}}
	\end{minipage} 
    \begin{minipage}[b]{121pt}
		\centering
		{\subcaption{\label{fig_demo_generate}}\includegraphics[width=121pt]{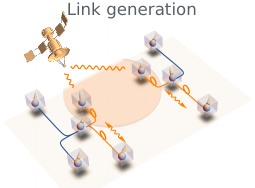}}
	\end{minipage} 
     \begin{minipage}[b]{243pt}
		\centering
		{\subcaption{\label{fig_resource}}\includegraphics[width=243pt]{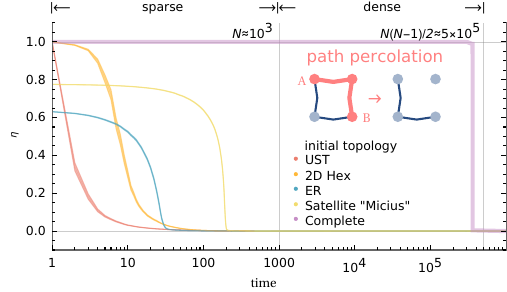}}
	\end{minipage} 
    \vspace{-3mm}
    
    \caption{Quantum communication network.
    \subref{fig_demo_remove}~Path removal. Communication of $1$ qubit costs $l$ ebits of entanglement, where $l$ is the path length between the two users.
    \subref{fig_demo_generate}~Link generation. Links are regenerated by optical fibers or satellites to replenish entanglement. 
    \subref{fig_resource}~The network availability $\eta$, i.e.,~the number of pairs that can still communicate, as a function of time. 
    Sparse initial network topologies, including uniform spanning tree (UST), two-dimensional honeycomb lattice (2D Hex), Erdős--Rényi (ER) (average degree $\left\langle k\right\rangle =2.0$), and a ``Micius''-satellite-based~\cite{micius_yllyczrcllshdlzlclwxwpep20}  synthetic network~\cite{q-netw_bccc21} (parameters: $N=10^3$, $R_\text{rec}=0.75$~m, $\eta=0.1$, $R=1.8\times 10^3$~km, and $n_p=50$ photons), disintegrate rapidly in fewer than $N$ time steps. On the contrary, a complete graph
    dismantles after $\sim N^2$ time steps.
    Each curve represents an average of $100$ samples. Error bands represent one standard error.
    \hfill\hfill}
\end{figure}

To address question (i), we observe that without link generation, realistically sparse quantum communication network topologies are highly susceptible to path percolation, causing them to fragment quickly. This vulnerability underscores the importance of link replenishment, leading us to question (ii): we find that the network topology can evolve into different phases, depending on the value of $\alpha$. For small $\alpha$, the network enters a disconnected ``DOWN'' phase, whereas for large $\alpha$, it undergoes a transition to a connected ``UP'' phase. This dynamic behavior leads to a phase diagram, where the transition boundary varies based on the physical implementation of link generation in the quantum communication network.

For instance, in an optical fiber implementation, the network topology is fixed. The fragility of quantum entanglement limits its establishment to $\sim 10^2$~km through optical fibers~\cite{q-netw-200km_dtyshhktnas09,*q-netw-fiber_imtat13}, allowing link generation only for neighboring pairs. Conversely, in satellite architectures, photon loss and turbulence predominantly occur in the lower $\sim 10$~km of the atmosphere~\cite{q-netw-satellite_ycllzrclldllgxllyjljrhzzwczlclspwp17}. Hence, links can be added with some probability between any pair of nodes 
at a much larger range, typically $\sim 10^3$~km~\cite{q-netw-satellite_ycllzrclldllgxllyjljrhzzwczlclspwp17}, as we discuss in more detail later.
Although the future structure of large-scale quantum networks is unknown, it will be of substantial complexity~\cite{q-netw_npb23,*Chepuri2023}, likely emerging from a synergy of technologies, combining satellite links with optical fibers~\cite{Orieux2016, *Razavi2018, q-netw-satellite_fpadglmmstdgktvdb23}.
Even with all the unknowns, it is reasonable to consider the case of allocating resources efficiently for link generation. Since adding links within a connected component does not increase the number of communicable user pairs, we strategically limit the generation of links to those that connect separate components, serving as a ``cross-linking'' scheme of link replenishment.

\begin{figure}[t!]
    \centering
    \begin{minipage}[b]{121pt}
		\centering
		{\subcaption{\label{fig_transition_alpha}}\includegraphics[width=121pt]{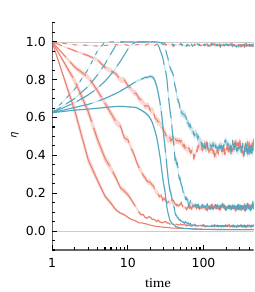}}
	\end{minipage} 
    \begin{minipage}[b]{121pt}
		\centering
		{\subcaption{\label{fig_transition}}\includegraphics[width=121pt]{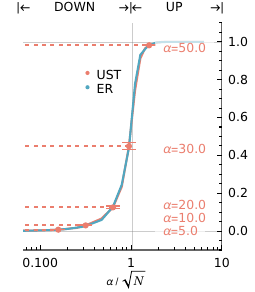}}
	\end{minipage} 
    \begin{minipage}[b]{121pt}
		\centering
		{\subcaption{\label{fig_transition_n_alpha}\includegraphics[width=115pt]{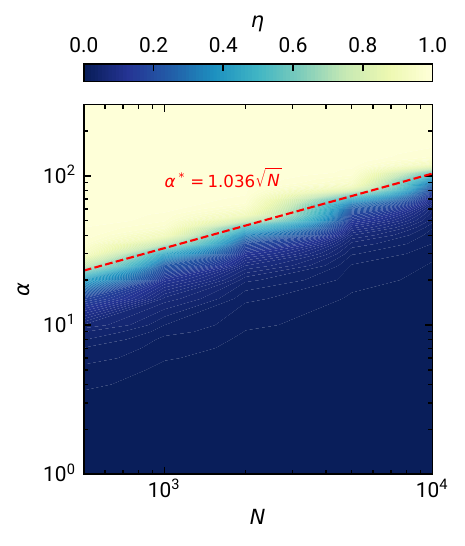}}}
	\end{minipage} 
    \begin{minipage}[b]{121pt}
		\centering
		{\subcaption{\label{fig_transition_general}}\includegraphics[width=121pt]{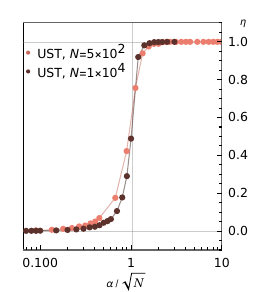}}
	\end{minipage} 

    \vspace{-3mm}
    \caption{``Cross-linking'' replenishment scheme. Only links connecting separate components are generated.
    \subref{fig_transition_alpha}~Steady states with both path removal and a link generation rate $\alpha>1$, starting from a uniform spanning tree (UST) initial topology ($N=10^3$). 
    \subref{fig_transition}~Variations in $\alpha$ result in distinct steady states. Each curve represents an average of $100$ samples. Error bands/bars represent one standard error. An ER initial topology  ($N=10^3$, $\left\langle k\right\rangle =2.0$) is also depicted, showing results identical to those from UST initial topology.
    \subref{fig_transition_n_alpha}~Phase diagram of path percolation using the cross-linking scheme. In the connected phase (yellow), the link generation rate adheres to 
    $\alpha^* \sim \sqrt{N}$
    (red dashed line).
    \subref{fig_transition_general}~Steady states for networks of increasing size $N$. As the size increases, the transition becomes sharper near the threshold, $\alpha^*/\sqrt{N}\approx 1.036$. 
    \hfill\hfill}
\end{figure}

Our key finding is that under the cross-linking scheme, the final steady-state network topology remains independent of the initial topology, whether in the ``DOWN'' or ``UP'' phase. The components consistently exhibit a random-tree-like structure, characterized by the scaling relation $l \sim \sqrt{s}$, where $s$ is the component size. Moreover, at the transition between the two phases the link generation rate  is found to scale
as $\alpha^* \sim \sqrt{N}$.

    \emph{Path percolation.---}We consider the following model comprising two steps: (i) Communication [Fig.~\ref{fig_demo_remove}]:~At each time step, a pair of nodes are uniformly randomly selected (communication rate $=1$~qubit/time step). If the pair cannot communicate, another pair is selected until a suitable pair is found. Then, a path connecting the two nodes is selected, with all participating links removed, marking a communication event. 
    To optimize resource usage, we select the path randomly among all
    \emph{shortest} paths connecting the two nodes.
    (ii) Link generation [Fig.~\ref{fig_demo_generate}]:~At each time step, $\alpha$ links are uniformly randomly selected, and sequentially placed for the set of node pairs where the two nodes are in different components of the network (link generation rate $=\alpha$~ebit/time step). 
We first consider the case $\alpha=0$. Note that $\alpha<1$ cannot lead to a nontrivial steady state, as in each step more links are being removed than generated.

As an order parameter to quantify the connectivity of the quantum network, we examine the \emph{network availability} $\eta$, given by the fraction of pairs of users that are connected and thus can communicate:
\begin{equation}
    \eta=\sum_{s=1}^{\infty} N(s)\frac{s(s-1)}{N(N-1)}=\frac{\left\langle s\right\rangle-1}{N-1}\,
\end{equation}
where $N(s)=N v(s)$ is the number of connected components of size $s$, and thus $v(s)$ the number of components \emph{per site}, such that $\sum_s s v(s)=1$. Here, $\left\langle s\right\rangle=\sum_s s^2 {v(s)}$ is the mean component size per site, a quantity often considered for Bernoulli percolation~\cite{percolation-theor_c02}. 
In Fig.~\ref{fig_resource} we show the network availability as
a function of a network of size $N$ across various initial network topologies.
In the unrealistic scenario of a complete graph of $N$ nodes, where all nodes were directly connected, there would be no deviation from the usual link removal scenario until the removal of longer paths becomes prevalent. Therefore, the complete graph falls apart at the point where $\sim N^2$ qubits of information have been communicated. In contrast, practical quantum satellite networks~\cite{q-netw_bccc21} already dismantle after communicating $<N$ qubits [Fig.~\ref{fig_resource}]. Moreover, even at the initial topology of the quantum satellite architecture, only $<80\%$ of the pairs can communicate, as not all users are part of the giant connected component. Sparser random networks, like Erd\H{o}s--R\'enyi (ER) random graphs with a finite average degree, $\left\langle k\right\rangle$, show qualitatively similar behavior. Two-dimensional network topologies---motivated by optical fiber-based two-dimensional grid-like architectures---also dismantle long before communicating $\sim N$ qubits. As the most extreme case of sparsity, tree-like quantum networks [e.g.,~uniform spanning trees (USTs)] become macroscopically fragmented already after communicating the very first qubit.

Once link generation with $\alpha>1$ is turned on, the steady state $\eta$ depends on $\alpha$ [Fig.~\ref{fig_transition_alpha}].
In general, we expect two phases in the steady state [Fig.~\ref{fig_transition}]: i) a disconnected ``DOWN'' phase with small $\alpha$ and $\eta\ll1$;
and ii) a connected ``UP'' phase, with large $\alpha$ and $1-\eta\ll 1$. Notably, both phases are found to be independent of the initial topology (e.g., UST vs.~ER).
By considering various network sizes $N$, we determine the transitional threshold $\alpha^*$ as a function of $N$, finding a power-law scaling $\alpha^* \sim \sqrt{N}$ [Fig.~\ref{fig_transition_n_alpha}].
Indeed, in Fig.~\ref{fig_transition_general}, we see that the $\eta$ vs.~$\alpha/\sqrt{N}$ curve becomes sharper as $N$ increases, intersecting at around $\alpha^*/\sqrt{N}\approx 1.036(10)$.

\begin{figure}[t!]
    \centering
    \begin{minipage}[b]{240pt}
		\centering
        {\subcaption{\label{fig_stationary_graph}}\includegraphics[width=240pt]{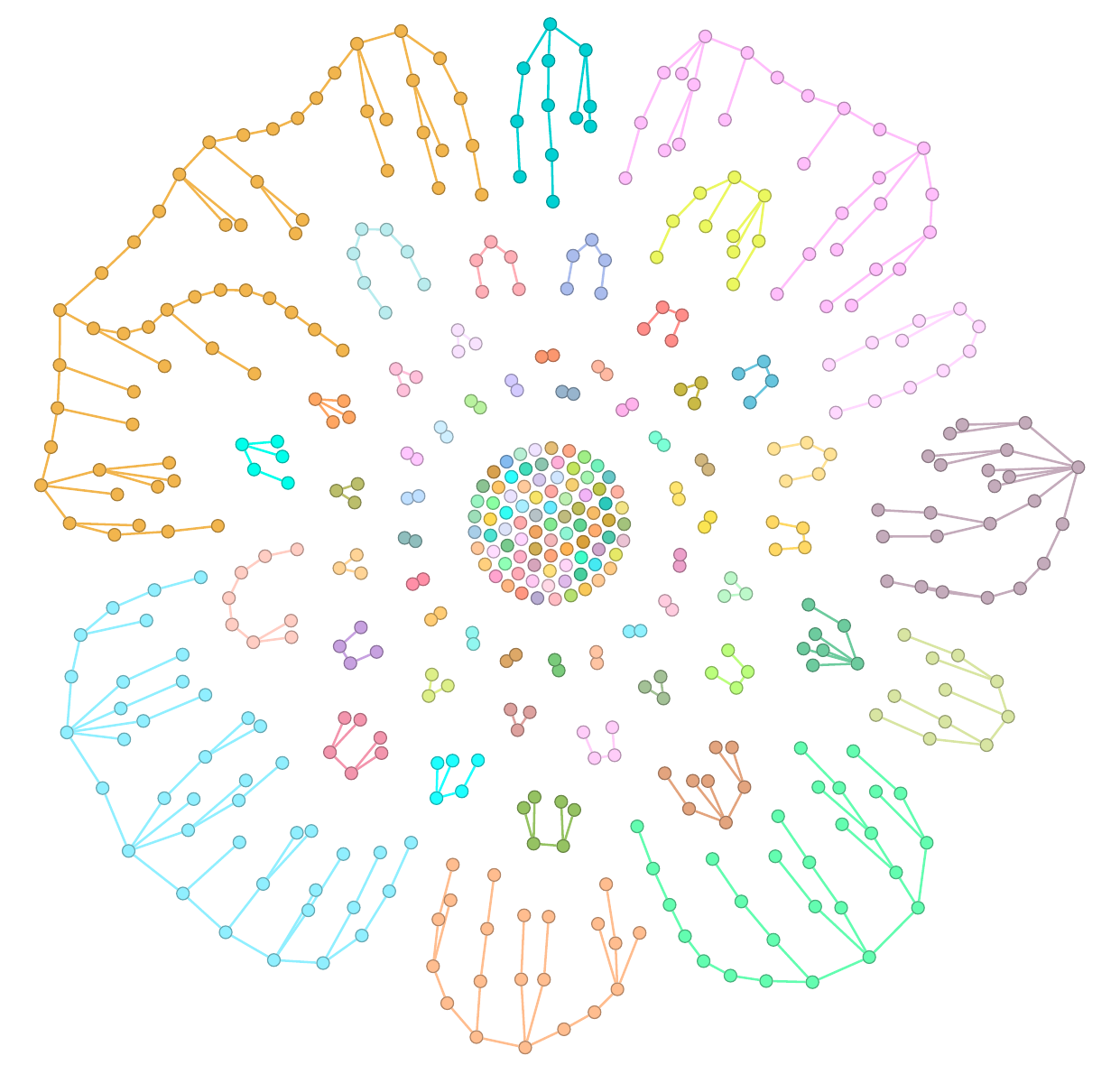}}
    \end{minipage} 
    \begin{minipage}[b]{110pt}
		\centering
        {\subcaption{\label{fig_component_path_length}}\includegraphics[width=110pt]{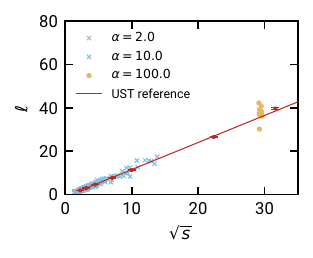}}
    \end{minipage} 
    \begin{minipage}[b]{120pt}
		\centering
        {\subcaption{\label{fig_stationary_path}}\includegraphics[width=120pt]{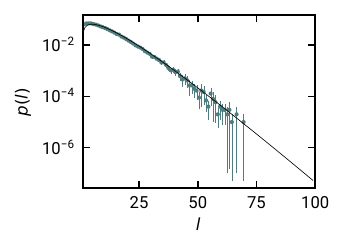}}
    \end{minipage} 
    \vspace{-3mm}
    \caption{ Tree-like components in the steady state, starting with an ER initial topology of $\alpha=10.0$.
    \subref{fig_stationary_graph}~Snapshot of the steady state. 
    \subref{fig_transition}~In the steady state  ($N=10^3$), all components exhibit tree-like behavior, i.e.,~the average characteristic path length $\ell$ scales as $\ell\sim\sqrt{s}$. This plot shows a collection of 10 samples for each $\alpha$.
    Error bars represent one standard error. 
    \subref{fig_stationary_path}~Length distribution of the paths removed in the steady state  ($N=10^3$). The solid line indicates theoretical prediction from the generalized Smoluchowski equation.
    \hfill\hfill}
\end{figure}

Our scheme of generating links exclusively between separate components ensures that the steady state is always a \emph{forest}, consisting only of tree-like components [Fig.~\ref{fig_stationary_graph}]. In the DOWN phase, typically more than $\alpha+1$ components exist,  allowing the addition of all $\alpha$ links in each step; while in the UP phase, the largest component size $s_{\mathrm{max}}$ is comparable to $N$, and thus it is not always possible to add all $\alpha$ links in each step.
The tree-like components lead to a key observation: for each component of size $s$, its typical path length $l$ is proportional to $ \sqrt{s}$ [Fig.~\ref{fig_component_path_length}]. 
Such a scaling is the characteristic of \emph{random trees}~\cite{random-tree_bck01}, describing many generic network topologies such as the UST.
This random tree assumption facilitates our theoretical analysis presented next.

\emph{Theoretical Analysis.---}In the disconnected DOWN phase, the size of the largest component, $s_{\max}$, is constrained by the parameter $\alpha$. This is because the length of the shortest paths is proportional to $\sqrt{s_{\max}}$, and the generation of links must compensate for the removal of an equivalent number of links, leading to $\alpha^* \sim\sqrt{s_{\max}}$. If we assume that the critical threshold is located at $s_{\max}\sim N$, it follows that $\alpha^* \sim\sqrt{N}$, aligning with our simulation results [Fig.~\ref{fig_transition_n_alpha}].

\begin{figure}[t!]
    \centering
    \begin{minipage}[b]{121pt}
		\centering
		{\subcaption{\label{fig_s_tree_small}}\includegraphics[width=121pt]{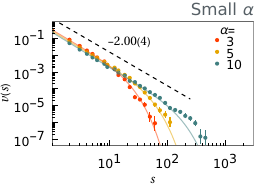}}
	\end{minipage} 
    \begin{minipage}[b]{121pt}
		\centering
		{\subcaption{\label{fig_s_tree_large}}\includegraphics[width=121pt]{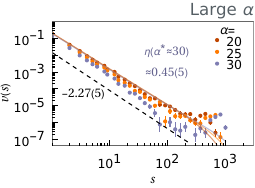}}
	\end{minipage}
    
    \begin{minipage}[b]{121pt}
		\centering
		{\subcaption{\label{fig_s_downlink}}\includegraphics[width=121pt]{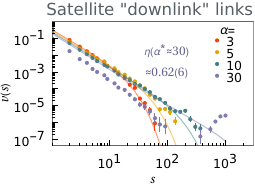}}
	\end{minipage}
    \begin{minipage}[b]{121pt}
		\centering
		{\subcaption{\label{fig_s_random}}\includegraphics[width=121pt]{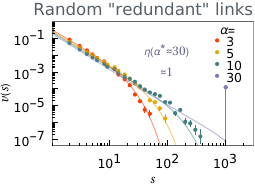}}
	\end{minipage} 
    
    \vspace{-3mm}
    
    \caption{\label{fig_s_tree}Steady-state component size distribution $v(s)$ compared between theoretical predictions (lines) and simulations (dots) with UST initial topology ($N=10^3$).
    \subref{fig_s_tree_small}~For $\alpha \ll \sqrt{N}$, the theoretical prediction matches simulation results, both following a power law $v(s)\sim s^{-2}$ with an exponential cutoff.
    \subref{fig_s_tree_large}~For $\alpha \sim \sqrt{N}$, the theoretical prediction fails to match $v(s)$ but correctly predicts the power law behavior $v(s)\sim s^{-9/4}$ near the critical threshold ($\alpha^*\approx 30$).
    \subref{fig_s_downlink}~A more realistic downlink link replenishment scheme, where peripheral nodes have less chance to connect~\cite{q-netw_bccc21}. For $\alpha \ll \sqrt{N}$, theoretical predictions remain accurate. However, at $\alpha= \alpha^*$, the downlink scheme achieves higher network availability ($\eta\approx0.62$) than the random cross-linking scheme ($\eta\approx0.45$).
    \subref{fig_s_random}~A random scheme that also allows adding ``redundant'' links within components, potentially forming shortcuts between users. Again, for $\alpha \ll \sqrt{N}$, theoretical predictions match the simulations; however, at $\alpha= \alpha^*$, the network availability increases to $\eta\approx 1$.
    \hfill\hfill}
\end{figure}

Following path percolation analytically is a challenging task, as it is a non-local process that depends sensitively on the network topology of each cluster.
Yet, at least in the disconnected phase---where clusters are typically small---we can describe the process using a generalized Smoluchowski equation~\cite{smoluchowski_s16}, which follows the change in the distribution of component sizes at each step. 
We find that the solution of $v(s)$ by the Smoluchowski equation (SI, Section~2) closely matches the simulation results [Fig.~\ref{fig_s_tree_small}] and only deviates when $\alpha \sim \sqrt{N}$, i.e.,~as the network transitions into the connected phase [Fig.~\ref{fig_s_tree_large}].
{From Fig.~\ref{fig_s_tree_small}, we further derive the distribution of the resource cost $l$ per step in the steady state of the DOWN phase, which is given by $p(l)\sim \sum_s s(s-1) v(s) p_\text{UST}(l|s)$
(SI,~Section~4). Here, $p_\text{UST}(l|s)$ is the path length distribution of a UST of size $s$  (SI,~Section~1). The theoretical prediction agrees with simulations [Fig.~\ref{fig_stationary_path}].}

In particular, at the critical threshold, the Smoluchowski solution is consistent with a power-law tail,
as long as $\tau<3$. With this ansatz, the Smoluchowski equation implies (SI, Section~3)
\begin{equation}
\label{eq_smoluchowski_k_alpha}
\alpha^*\simeq \frac{3-\tau }{7-2 \tau } \zeta\left(\frac{3}{2}\right)\sqrt{N},
\end{equation}
where $\zeta(s)$ is the Riemann zeta function. As a key result, by considering
the fourth moment of $v(s)$, we find that the critical exponent $\tau$ is initially close to $2$ when $\alpha$ is small ($\sim N^0$). Yet, as $\alpha\to N^{1/2}$, it approaches $\tau=9/4$.
The theoretical predictions for $\tau$ are consistent with the simulation results in both regimes [Figs.~\ref{fig_s_tree_small}~and~\ref{fig_s_tree_large}].
Note, however, that substituting $\tau=9/4$ into Eq.~\eqref{eq_smoluchowski_k_alpha} yields $\alpha^*\simeq 0.783 \sqrt{N}$, slightly off the simulation data in Fig.~\ref{fig_transition_n_alpha}. This is expected, as fluctuations (including rare large clusters and their topology) become important at the transition point.

\emph{Further link replenishment schemes.---}We have shown that, within the ``cross-linking'' scheme, the steady state $v(s)$ is insensitive to initial conditions.
However,  $v(s)$ might still depend sensitively on different link replenishment schemes.  As an alternative, we first explore a more realistic ``downlink'' scheme that utilizes satellite-to-ground downlink channels, taking into account turbulent contributions and trajectory inclination~\cite{qkd-satellite_btdnv09}. We model the service area as a territorial disk-shaped region directly below the satellite~\cite{q-netw_bccc21}. Nodes are randomly distributed within this disk. Akin to the cross-linking scheme, the probability $\Pi_{ij}$ of establishing a link between two nodes $i$ and $j$ is nonzero only if the two nodes are from separate components, in which case it is given by $\Pi_{ij}=1-\left(1-p_i p_j\right)^{n_p}$, with $n_p$ set at $50$ photons for link attempts~\cite{q-netw_bccc21}. The probability $p_i$ ($p_j$) that a photon from the satellite reaches a node depends on the node's distance from the disk’s center, with closer nodes having a higher probability of establishing a link~\cite{q-netw_bccc21}.

As Fig.~\ref{fig_s_downlink} shows, when $\alpha\ll \sqrt{N}$, $v(s)$ adheres to the theory, implying the generality of the Smoluchowski dynamics beyond purely random link generation. However, at $\alpha\sim \sqrt{N}$, the network availability $\eta$ under the downlink scheme is higher than random link generation [Fig.~\ref{fig_s_tree_large}]. This indicates that while satellite downlinks may be less effective at connecting peripheral nodes, they provide significant connectivity advantages for nodes near the center, offsetting the peripheral connectivity limitations.

Another possible ``redundancy'' scheme is to allow adding links within the same component, i.e.,~allow loops in the network, as these redundant loops could potentially reduce the path length $l$ and thus the resource cost.
While this feature is uncommon for optical fiber connections, it may occur through satellite communication. As Fig.~\ref{fig_s_random} shows, adding loops within components does not alter the component size distribution when $\alpha\ll \sqrt{N}$, implying that the components may still be considered tree-like.
However, at $\alpha\sim \sqrt{N}$, 
incorporating loops is found to greatly enhance the network availability $\eta$  [Fig.~\ref{fig_s_random}], compared to the cross-linking scheme.

\emph{Discussion.---}Communication dynamics in quantum networks fundamentally differs from classical communication due to the presence of a unique physical resource layer based on entanglement links, which are consumed during communication. This is contrary to classical communication, where encoding in classical electromagnetic signals already encapsulates the information, without the need for an additional layer~\cite{q-netw-satellite_fpadglmmstdgktvdb23}. The utilization of this quantum-specific physical layer initiates a non-local path percolation process that results in novel phenomena.
We have shown that generating new links with a finite rate $\alpha$ leads to non-trivial, yet often inherently disconnected steady states below a threshold $\alpha^*\sim\sqrt{N}$ in a network of $N$ nodes. 
When generating links exclusively between separate components,
the steady state is independent of the initial network topology and is found to be a collection of tree components that closely resemble uniform spanning trees (USTs).
As a consequence, our results indicate that in the steady state the resulting network components of size $s$ are not small-world, but rather follow UST-like statistics (also seen in 2D), with an average path length of $l~\sim\sqrt{s}$.
Counterintuitively, adding links within the same component is found to boost network availability for large enough values of $\alpha$.
These findings emphasize the importance of strategic planning in future quantum link generation schemes---potentially drawing on the extensive literature on classical link addition/recovery~\cite{link-add_chdz13,*link-add_czqmg22,*interdepend-recover_wmdctb22,*link-add_msmhp23}, but tailored to the new path percolation context.

Our work can be extended in several directions. Already at $\alpha=0$, exploring the nature of the phase transition of how the network dismantles and the underlying universality classes (if any) is an exciting problem even on low-dimensional lattices. Similarly, further studies are needed to understand the transition between the connected and disconnected steady states as a function of $\alpha$.
Note that just like in the case of the classical Internet, future quantum communication protocols might not strictly operate through shortest paths~\cite{q-netw-long-path_hdlcftghm24}, but even so shortest paths serve as an informative best-case scenario. Exploring the impact of less efficient routing protocols, that only use local information, is another open challenge.
Our model can be also extended by considering alternative strategies for link generation. For example, motivated by optical fiber infrastructures, we could add links uniformly randomly from the list of removed links from the initial configuration. Such changes in the process are expected to lead to qualitatively different behavior.

In future applications, user engagement might also be highly heterogeneous, as it is likely that instead of randomly communicating with other users, most users would typically want to connect to their critical utilities, like banks and other service providers.
Our proposed framework can be generalized for arbitrary heterogeneities in user behavior.
Such situations also call for policies to control which users are allowed to communicate at any time, or how many requests a central hub can satisfy.

\emph{Note.---}Upon finalizing our manuscript, we became aware of Ref.~\cite{kim2024}, which introduces shortest path percolation motivated by classical network problems, presenting detailed results for Erd\H{o}s--R\'enyi random graphs in the lack of adding links, $\alpha=0$. Note that $\alpha>0$ is expected to be relevant also in the related classical problems.
One example could be adding intercontinental flights to connect fragmented components (continents) in airline networks, where path percolation is expected to happen~\cite{kim2024}.

\emph{Acknowledgment.---}We thank Ravi T.~C.~Chepuri, Helen S. Ansell, Chana Lyubich, Ruiting (Grace) Xie, Dániel Keliger, Ágnes
Kúsz, Pierfrancesco Dionigi and Miklós Abért for helpful comments and discussion.
This work greatly benefited from the 2023 Focused Workshop on `Networks and Their Limits' held at the Erdős Center (part of the Alfréd Rényi Institute of Mathematics) in Budapest, Hungary.
The workshop was supported by the ERC Synergy grant DYNASNET 810115.
The work of IAK and XM was supported by the National Science Foundation under Grant No.~PHY-2310706 of the QIS program in the Division of Physics.

\bibliography{refs}

\begin{thebibliography}{43}%
\makeatletter
\providecommand \@ifxundefined [1]{%
 \@ifx{#1\undefined}
}%
\providecommand \@ifnum [1]{%
 \ifnum #1\expandafter \@firstoftwo
 \else \expandafter \@secondoftwo
 \fi
}%
\providecommand \@ifx [1]{%
 \ifx #1\expandafter \@firstoftwo
 \else \expandafter \@secondoftwo
 \fi
}%
\providecommand \natexlab [1]{#1}%
\providecommand \enquote  [1]{``#1''}%
\providecommand \bibnamefont  [1]{#1}%
\providecommand \bibfnamefont [1]{#1}%
\providecommand \citenamefont [1]{#1}%
\providecommand \href@noop [0]{\@secondoftwo}%
\providecommand \href [0]{\begingroup \@sanitize@url \@href}%
\providecommand \@href[1]{\@@startlink{#1}\@@href}%
\providecommand \@@href[1]{\endgroup#1\@@endlink}%
\providecommand \@sanitize@url [0]{\catcode `\\12\catcode `\$12\catcode `\&12\catcode `\#12\catcode `\^12\catcode `\_12\catcode `\%12\relax}%
\providecommand \@@startlink[1]{}%
\providecommand \@@endlink[0]{}%
\providecommand \url  [0]{\begingroup\@sanitize@url \@url }%
\providecommand \@url [1]{\endgroup\@href {#1}{\urlprefix }}%
\providecommand \urlprefix  [0]{URL }%
\providecommand \Eprint [0]{\href }%
\providecommand \doibase [0]{https://doi.org/}%
\providecommand \selectlanguage [0]{\@gobble}%
\providecommand \bibinfo  [0]{\@secondoftwo}%
\providecommand \bibfield  [0]{\@secondoftwo}%
\providecommand \translation [1]{[#1]}%
\providecommand \BibitemOpen [0]{}%
\providecommand \bibitemStop [0]{}%
\providecommand \bibitemNoStop [0]{.\EOS\space}%
\providecommand \EOS [0]{\spacefactor3000\relax}%
\providecommand \BibitemShut  [1]{\csname bibitem#1\endcsname}%
\let\auto@bib@innerbib\@empty
\bibitem [{\citenamefont {Kimble}(2008)}]{q-internet_k08}%
  \BibitemOpen
  \bibfield  {author} {\bibinfo {author} {\bibfnamefont {H.~J.}\ \bibnamefont {Kimble}},\ }\bibfield  {title} {\bibinfo {title} {The quantum internet},\ }\href {https://doi.org/10.1038/nature07127} {\bibfield  {journal} {\bibinfo  {journal} {Nature}\ }\textbf {\bibinfo {volume} {453}},\ \bibinfo {pages} {1023} (\bibinfo {year} {2008})}\BibitemShut {NoStop}%
\bibitem [{\citenamefont {Wehner}\ \emph {et~al.}(2018)\citenamefont {Wehner}, \citenamefont {Elkouss},\ and\ \citenamefont {Hanson}}]{q-internet_weh18}%
  \BibitemOpen
  \bibfield  {author} {\bibinfo {author} {\bibfnamefont {S.}~\bibnamefont {Wehner}}, \bibinfo {author} {\bibfnamefont {D.}~\bibnamefont {Elkouss}},\ and\ \bibinfo {author} {\bibfnamefont {R.}~\bibnamefont {Hanson}},\ }\bibfield  {title} {\bibinfo {title} {Quantum internet: {{A}} vision for the road ahead},\ }\href {https://doi.org/10.1126/science.aam9288} {\bibfield  {journal} {\bibinfo  {journal} {Science}\ }\textbf {\bibinfo {volume} {362}},\ \bibinfo {pages} {eaam9288} (\bibinfo {year} {2018})}\BibitemShut {NoStop}%
\bibitem [{\citenamefont {Gisin}\ \emph {et~al.}(2002)\citenamefont {Gisin}, \citenamefont {Ribordy}, \citenamefont {Tittel},\ and\ \citenamefont {Zbinden}}]{q-crypto_grtz02}%
  \BibitemOpen
  \bibfield  {author} {\bibinfo {author} {\bibfnamefont {N.}~\bibnamefont {Gisin}}, \bibinfo {author} {\bibfnamefont {G.~G.}\ \bibnamefont {Ribordy}}, \bibinfo {author} {\bibfnamefont {W.}~\bibnamefont {Tittel}},\ and\ \bibinfo {author} {\bibfnamefont {H.}~\bibnamefont {Zbinden}},\ }\bibfield  {title} {\bibinfo {title} {Quantum cryptography},\ }\href {https://doi.org/10.1103/RevModPhys.74.145} {\bibfield  {journal} {\bibinfo  {journal} {Reviews of Modern Physics}\ }\textbf {\bibinfo {volume} {74}},\ \bibinfo {pages} {145} (\bibinfo {year} {2002})}\BibitemShut {NoStop}%
\bibitem [{\citenamefont {Bennett}\ and\ \citenamefont {Wiesner}(1992)}]{q-superdense_bw92}%
  \BibitemOpen
  \bibfield  {author} {\bibinfo {author} {\bibfnamefont {C.~H.}\ \bibnamefont {Bennett}}\ and\ \bibinfo {author} {\bibfnamefont {S.~J.}\ \bibnamefont {Wiesner}},\ }\bibfield  {title} {\bibinfo {title} {Communication via {{One-}} and {{Two-Particle Operators}} on {{Einstein-Podolsky-Rosen States}}},\ }\href {https://doi.org/10.1103/PhysRevLett.69.2881} {\bibfield  {journal} {\bibinfo  {journal} {Physical Review Letters}\ }\textbf {\bibinfo {volume} {69}},\ \bibinfo {pages} {2881} (\bibinfo {year} {1992})}\BibitemShut {NoStop}%
\bibitem [{\citenamefont {Dumitrescu}\ \emph {et~al.}(2018)\citenamefont {Dumitrescu}, \citenamefont {McCaskey}, \citenamefont {Hagen}, \citenamefont {Jansen}, \citenamefont {Morris}, \citenamefont {Papenbrock}, \citenamefont {Pooser}, \citenamefont {Dean},\ and\ \citenamefont {Lougovski}}]{Dumitrescu2018}%
  \BibitemOpen
  \bibfield  {author} {\bibinfo {author} {\bibfnamefont {E.}~\bibnamefont {Dumitrescu}}, \bibinfo {author} {\bibfnamefont {A.}~\bibnamefont {McCaskey}}, \bibinfo {author} {\bibfnamefont {G.}~\bibnamefont {Hagen}}, \bibinfo {author} {\bibfnamefont {G.}~\bibnamefont {Jansen}}, \bibinfo {author} {\bibfnamefont {T.}~\bibnamefont {Morris}}, \bibinfo {author} {\bibfnamefont {T.}~\bibnamefont {Papenbrock}}, \bibinfo {author} {\bibfnamefont {R.}~\bibnamefont {Pooser}}, \bibinfo {author} {\bibfnamefont {D.}~\bibnamefont {Dean}},\ and\ \bibinfo {author} {\bibfnamefont {P.}~\bibnamefont {Lougovski}},\ }\bibfield  {title} {\bibinfo {title} {Cloud quantum computing of an atomic nucleus},\ }\bibfield  {journal} {\bibinfo  {journal} {Physical Review Letters}\ }\textbf {\bibinfo {volume} {120}},\ \href {https://doi.org/10.1103/physrevlett.120.210501} {10.1103/physrevlett.120.210501} (\bibinfo {year} {2018})\BibitemShut {NoStop}%
\bibitem [{\citenamefont {Gisin}\ and\ \citenamefont {Thew}(2007)}]{q-commun_gt07}%
  \BibitemOpen
  \bibfield  {author} {\bibinfo {author} {\bibfnamefont {N.}~\bibnamefont {Gisin}}\ and\ \bibinfo {author} {\bibfnamefont {R.}~\bibnamefont {Thew}},\ }\bibfield  {title} {\bibinfo {title} {Quantum communication},\ }\href {https://doi.org/10.1038/nphoton.2007.22} {\bibfield  {journal} {\bibinfo  {journal} {Nature Photonics}\ }\textbf {\bibinfo {volume} {1}},\ \bibinfo {pages} {165} (\bibinfo {year} {2007})}\BibitemShut {NoStop}%
\bibitem [{\citenamefont {Chitambar}\ and\ \citenamefont {Gour}(2019)}]{q-resour_cg19}%
  \BibitemOpen
  \bibfield  {author} {\bibinfo {author} {\bibfnamefont {E.}~\bibnamefont {Chitambar}}\ and\ \bibinfo {author} {\bibfnamefont {G.}~\bibnamefont {Gour}},\ }\bibfield  {title} {\bibinfo {title} {Quantum resource theories},\ }\href {https://doi.org/10.1103/RevModPhys.91.025001} {\bibfield  {journal} {\bibinfo  {journal} {Reviews of Modern Physics}\ }\textbf {\bibinfo {volume} {91}},\ \bibinfo {pages} {025001} (\bibinfo {year} {2019})}\BibitemShut {NoStop}%
\bibitem [{\citenamefont {Meng}\ \emph {et~al.}(2024)\citenamefont {Meng}, \citenamefont {Piparo}, \citenamefont {Nemoto},\ and\ \citenamefont {Kovács}}]{meng2024quantum}%
  \BibitemOpen
  \bibfield  {author} {\bibinfo {author} {\bibfnamefont {X.}~\bibnamefont {Meng}}, \bibinfo {author} {\bibfnamefont {N.~L.}\ \bibnamefont {Piparo}}, \bibinfo {author} {\bibfnamefont {K.}~\bibnamefont {Nemoto}},\ and\ \bibinfo {author} {\bibfnamefont {I.~A.}\ \bibnamefont {Kovács}},\ }\href@noop {} {\bibinfo {title} {Quantum networks enhanced by distributed quantum memories}} (\bibinfo {year} {2024}),\ \Eprint {https://arxiv.org/abs/2403.16367} {arXiv:2403.16367 [quant-ph]} \BibitemShut {NoStop}%
\bibitem [{\citenamefont {Kim}\ and\ \citenamefont {Lee}(2021)}]{q-resour_kl21}%
  \BibitemOpen
  \bibfield  {author} {\bibinfo {author} {\bibfnamefont {H.-J.}\ \bibnamefont {Kim}}\ and\ \bibinfo {author} {\bibfnamefont {S.}~\bibnamefont {Lee}},\ }\bibfield  {title} {\bibinfo {title} {One-shot static entanglement cost of bipartite quantum channels},\ }\href {https://doi.org/10.1103/PhysRevA.103.062415} {\bibfield  {journal} {\bibinfo  {journal} {Physical Review A}\ }\textbf {\bibinfo {volume} {103}},\ \bibinfo {pages} {062415} (\bibinfo {year} {2021})}\BibitemShut {NoStop}%
\bibitem [{\citenamefont {Ghavasieh}\ and\ \citenamefont {De~Domenico}(2024)}]{netw-sparse_gdd24}%
  \BibitemOpen
  \bibfield  {author} {\bibinfo {author} {\bibfnamefont {A.}~\bibnamefont {Ghavasieh}}\ and\ \bibinfo {author} {\bibfnamefont {M.}~\bibnamefont {De~Domenico}},\ }\bibfield  {title} {\bibinfo {title} {Diversity of information pathways drives sparsity in real-world networks},\ }\href {https://doi.org/10.1038/s41567-023-02330-x} {\bibfield  {journal} {\bibinfo  {journal} {Nature Physics}\ }\textbf {\bibinfo {volume} {20}},\ \bibinfo {pages} {512} (\bibinfo {year} {2024})}\BibitemShut {NoStop}%
\bibitem [{\citenamefont {{\.Z}ukowski}\ \emph {et~al.}(1993)\citenamefont {{\.Z}ukowski}, \citenamefont {Zeilinger}, \citenamefont {Horne},\ and\ \citenamefont {Ekert}}]{entangle-swap_zzhe93}%
  \BibitemOpen
  \bibfield  {author} {\bibinfo {author} {\bibfnamefont {M.}~\bibnamefont {{\.Z}ukowski}}, \bibinfo {author} {\bibfnamefont {A.}~\bibnamefont {Zeilinger}}, \bibinfo {author} {\bibfnamefont {M.~A.}\ \bibnamefont {Horne}},\ and\ \bibinfo {author} {\bibfnamefont {A.~K.}\ \bibnamefont {Ekert}},\ }\bibfield  {title} {\bibinfo {title} {``{{Event-Ready-Detectors}}'' {{Bell Experiment}} via {{Entanglement Swapping}}},\ }\href {https://doi.org/10.1103/PhysRevLett.71.4287} {\bibfield  {journal} {\bibinfo  {journal} {Physical Review Letters}\ }\textbf {\bibinfo {volume} {71}},\ \bibinfo {pages} {4287} (\bibinfo {year} {1993})}\BibitemShut {NoStop}%
\bibitem [{\citenamefont {Riebe}\ \emph {et~al.}(2008)\citenamefont {Riebe}, \citenamefont {Monz}, \citenamefont {Kim}, \citenamefont {Villar}, \citenamefont {Schindler}, \citenamefont {Chwalla}, \citenamefont {Hennrich},\ and\ \citenamefont {Blatt}}]{entangle-swap_rmkvschb08}%
  \BibitemOpen
  \bibfield  {author} {\bibinfo {author} {\bibfnamefont {M.}~\bibnamefont {Riebe}}, \bibinfo {author} {\bibfnamefont {T.}~\bibnamefont {Monz}}, \bibinfo {author} {\bibfnamefont {K.}~\bibnamefont {Kim}}, \bibinfo {author} {\bibfnamefont {A.~S.}\ \bibnamefont {Villar}}, \bibinfo {author} {\bibfnamefont {P.}~\bibnamefont {Schindler}}, \bibinfo {author} {\bibfnamefont {M.}~\bibnamefont {Chwalla}}, \bibinfo {author} {\bibfnamefont {M.}~\bibnamefont {Hennrich}},\ and\ \bibinfo {author} {\bibfnamefont {R.}~\bibnamefont {Blatt}},\ }\bibfield  {title} {\bibinfo {title} {Deterministic entanglement swapping with an ion-trap quantum computer},\ }\href {https://doi.org/10.1038/nphys1107} {\bibfield  {journal} {\bibinfo  {journal} {Nature Physics}\ }\textbf {\bibinfo {volume} {4}},\ \bibinfo {pages} {839} (\bibinfo {year} {2008})}\BibitemShut {NoStop}%
\bibitem [{\citenamefont {Cohen}\ \emph {et~al.}(2000)\citenamefont {Cohen}, \citenamefont {Erez}, \citenamefont {{ben-Avraham}},\ and\ \citenamefont {Havlin}}]{netw-percolation_ceah00}%
  \BibitemOpen
  \bibfield  {author} {\bibinfo {author} {\bibfnamefont {R.}~\bibnamefont {Cohen}}, \bibinfo {author} {\bibfnamefont {K.}~\bibnamefont {Erez}}, \bibinfo {author} {\bibfnamefont {D.}~\bibnamefont {{ben-Avraham}}},\ and\ \bibinfo {author} {\bibfnamefont {S.}~\bibnamefont {Havlin}},\ }\bibfield  {title} {\bibinfo {title} {Resilience of the {{Internet}} to {{Random Breakdowns}}},\ }\href {https://doi.org/10.1103/PhysRevLett.85.4626} {\bibfield  {journal} {\bibinfo  {journal} {Physical Review Letters}\ }\textbf {\bibinfo {volume} {85}},\ \bibinfo {pages} {4626} (\bibinfo {year} {2000})}\BibitemShut {NoStop}%
\bibitem [{\citenamefont {Holme}\ \emph {et~al.}(2002)\citenamefont {Holme}, \citenamefont {Kim}, \citenamefont {Yoon},\ and\ \citenamefont {Han}}]{netw-attack_hkyh02}%
  \BibitemOpen
  \bibfield  {author} {\bibinfo {author} {\bibfnamefont {P.}~\bibnamefont {Holme}}, \bibinfo {author} {\bibfnamefont {B.~J.}\ \bibnamefont {Kim}}, \bibinfo {author} {\bibfnamefont {C.~N.}\ \bibnamefont {Yoon}},\ and\ \bibinfo {author} {\bibfnamefont {S.~K.}\ \bibnamefont {Han}},\ }\bibfield  {title} {\bibinfo {title} {Attack vulnerability of complex networks},\ }\href {https://doi.org/10.1103/PhysRevE.65.056109} {\bibfield  {journal} {\bibinfo  {journal} {Physical Review E}\ }\textbf {\bibinfo {volume} {65}},\ \bibinfo {pages} {056109} (\bibinfo {year} {2002})}\BibitemShut {NoStop}%
\bibitem [{\citenamefont {Yin}\ \emph {et~al.}(2020)\citenamefont {Yin}, \citenamefont {Li}, \citenamefont {Liao}, \citenamefont {Yang}, \citenamefont {Cao}, \citenamefont {Zhang}, \citenamefont {Ren}, \citenamefont {Cai}, \citenamefont {Liu}, \citenamefont {Li}, \citenamefont {Shu}, \citenamefont {Huang}, \citenamefont {Deng}, \citenamefont {Li}, \citenamefont {Zhang}, \citenamefont {Liu}, \citenamefont {Chen}, \citenamefont {Lu}, \citenamefont {Wang}, \citenamefont {Xu}, \citenamefont {Wang}, \citenamefont {Peng}, \citenamefont {Ekert},\ and\ \citenamefont {Pan}}]{micius_yllyczrcllshdlzlclwxwpep20}%
  \BibitemOpen
  \bibfield  {author} {\bibinfo {author} {\bibfnamefont {J.}~\bibnamefont {Yin}}, \bibinfo {author} {\bibfnamefont {Y.-H.}\ \bibnamefont {Li}}, \bibinfo {author} {\bibfnamefont {S.-K.}\ \bibnamefont {Liao}}, \bibinfo {author} {\bibfnamefont {M.}~\bibnamefont {Yang}}, \bibinfo {author} {\bibfnamefont {Y.}~\bibnamefont {Cao}}, \bibinfo {author} {\bibfnamefont {L.}~\bibnamefont {Zhang}}, \bibinfo {author} {\bibfnamefont {J.-G.}\ \bibnamefont {Ren}}, \bibinfo {author} {\bibfnamefont {W.-Q.}\ \bibnamefont {Cai}}, \bibinfo {author} {\bibfnamefont {W.-Y.}\ \bibnamefont {Liu}}, \bibinfo {author} {\bibfnamefont {S.-L.}\ \bibnamefont {Li}}, \bibinfo {author} {\bibfnamefont {R.}~\bibnamefont {Shu}}, \bibinfo {author} {\bibfnamefont {Y.-M.}\ \bibnamefont {Huang}}, \bibinfo {author} {\bibfnamefont {L.}~\bibnamefont {Deng}}, \bibinfo {author} {\bibfnamefont {L.}~\bibnamefont {Li}}, \bibinfo {author} {\bibfnamefont {Q.}~\bibnamefont {Zhang}}, \bibinfo {author} {\bibfnamefont {N.-L.}\ \bibnamefont {Liu}}, \bibinfo {author}
  {\bibfnamefont {Y.-A.}\ \bibnamefont {Chen}}, \bibinfo {author} {\bibfnamefont {C.-Y.}\ \bibnamefont {Lu}}, \bibinfo {author} {\bibfnamefont {X.-B.}\ \bibnamefont {Wang}}, \bibinfo {author} {\bibfnamefont {F.}~\bibnamefont {Xu}}, \bibinfo {author} {\bibfnamefont {J.-Y.}\ \bibnamefont {Wang}}, \bibinfo {author} {\bibfnamefont {C.-Z.}\ \bibnamefont {Peng}}, \bibinfo {author} {\bibfnamefont {A.~K.}\ \bibnamefont {Ekert}},\ and\ \bibinfo {author} {\bibfnamefont {J.-W.}\ \bibnamefont {Pan}},\ }\bibfield  {title} {\bibinfo {title} {Entanglement-based secure quantum cryptography over 1,120 kilometres},\ }\href {https://doi.org/10.1038/s41586-020-2401-y} {\bibfield  {journal} {\bibinfo  {journal} {Nature}\ }\textbf {\bibinfo {volume} {582}},\ \bibinfo {pages} {501} (\bibinfo {year} {2020})}\BibitemShut {NoStop}%
\bibitem [{\citenamefont {Brito}\ \emph {et~al.}(2021)\citenamefont {Brito}, \citenamefont {Canabarro}, \citenamefont {Cavalcanti},\ and\ \citenamefont {Chaves}}]{q-netw_bccc21}%
  \BibitemOpen
  \bibfield  {author} {\bibinfo {author} {\bibfnamefont {S.}~\bibnamefont {Brito}}, \bibinfo {author} {\bibfnamefont {A.}~\bibnamefont {Canabarro}}, \bibinfo {author} {\bibfnamefont {D.}~\bibnamefont {Cavalcanti}},\ and\ \bibinfo {author} {\bibfnamefont {R.}~\bibnamefont {Chaves}},\ }\bibfield  {title} {\bibinfo {title} {Satellite-{{Based Photonic Quantum Networks Are Small-World}}},\ }\href {https://doi.org/10.1103/PRXQuantum.2.010304} {\bibfield  {journal} {\bibinfo  {journal} {PRX Quantum}\ }\textbf {\bibinfo {volume} {2}},\ \bibinfo {pages} {010304} (\bibinfo {year} {2021})}\BibitemShut {NoStop}%
\bibitem [{\citenamefont {Dynes}\ \emph {et~al.}(2009)\citenamefont {Dynes}, \citenamefont {Takesue}, \citenamefont {Yuan}, \citenamefont {Sharpe}, \citenamefont {Harada}, \citenamefont {Honjo}, \citenamefont {Kamada}, \citenamefont {Tadanaga}, \citenamefont {Nishida}, \citenamefont {Asobe},\ and\ \citenamefont {Shields}}]{q-netw-200km_dtyshhktnas09}%
  \BibitemOpen
  \bibfield  {author} {\bibinfo {author} {\bibfnamefont {J.~F.}\ \bibnamefont {Dynes}}, \bibinfo {author} {\bibfnamefont {H.}~\bibnamefont {Takesue}}, \bibinfo {author} {\bibfnamefont {Z.~L.}\ \bibnamefont {Yuan}}, \bibinfo {author} {\bibfnamefont {A.~W.}\ \bibnamefont {Sharpe}}, \bibinfo {author} {\bibfnamefont {K.}~\bibnamefont {Harada}}, \bibinfo {author} {\bibfnamefont {T.}~\bibnamefont {Honjo}}, \bibinfo {author} {\bibfnamefont {H.}~\bibnamefont {Kamada}}, \bibinfo {author} {\bibfnamefont {O.}~\bibnamefont {Tadanaga}}, \bibinfo {author} {\bibfnamefont {Y.}~\bibnamefont {Nishida}}, \bibinfo {author} {\bibfnamefont {M.}~\bibnamefont {Asobe}},\ and\ \bibinfo {author} {\bibfnamefont {A.~J.}\ \bibnamefont {Shields}},\ }\bibfield  {title} {\bibinfo {title} {Efficient entanglement distribution over 200 kilometers},\ }\href {https://doi.org/10.1364/OE.17.011440} {\bibfield  {journal} {\bibinfo  {journal} {Optics Express}\ }\textbf {\bibinfo {volume} {17}},\ \bibinfo {pages} {11440} (\bibinfo {year}
  {2009})}\BibitemShut {NoStop}%
\bibitem [{\citenamefont {Inagaki}\ \emph {et~al.}(2013)\citenamefont {Inagaki}, \citenamefont {Matsuda}, \citenamefont {Tadanaga}, \citenamefont {Asobe},\ and\ \citenamefont {Takesue}}]{q-netw-fiber_imtat13}%
  \BibitemOpen
  \bibfield  {author} {\bibinfo {author} {\bibfnamefont {T.}~\bibnamefont {Inagaki}}, \bibinfo {author} {\bibfnamefont {N.}~\bibnamefont {Matsuda}}, \bibinfo {author} {\bibfnamefont {O.}~\bibnamefont {Tadanaga}}, \bibinfo {author} {\bibfnamefont {M.}~\bibnamefont {Asobe}},\ and\ \bibinfo {author} {\bibfnamefont {H.}~\bibnamefont {Takesue}},\ }\bibfield  {title} {\bibinfo {title} {Entanglement distribution over 300 km of fiber},\ }\href {https://doi.org/10.1364/OE.21.023241} {\bibfield  {journal} {\bibinfo  {journal} {Optics Express}\ }\textbf {\bibinfo {volume} {21}},\ \bibinfo {pages} {23241} (\bibinfo {year} {2013})}\BibitemShut {NoStop}%
\bibitem [{\citenamefont {Yin}\ \emph {et~al.}(2017)\citenamefont {Yin}, \citenamefont {Cao}, \citenamefont {Li}, \citenamefont {Liao}, \citenamefont {Zhang}, \citenamefont {Ren}, \citenamefont {Cai}, \citenamefont {Liu}, \citenamefont {Li}, \citenamefont {Dai}, \citenamefont {Li}, \citenamefont {Lu}, \citenamefont {Gong}, \citenamefont {Xu}, \citenamefont {Li}, \citenamefont {Li}, \citenamefont {Yin}, \citenamefont {Jiang}, \citenamefont {Li}, \citenamefont {Jia}, \citenamefont {Ren}, \citenamefont {He}, \citenamefont {Zhou}, \citenamefont {Zhang}, \citenamefont {Wang}, \citenamefont {Chang}, \citenamefont {Zhu}, \citenamefont {Liu}, \citenamefont {Chen}, \citenamefont {Lu}, \citenamefont {Shu}, \citenamefont {Peng}, \citenamefont {Wang},\ and\ \citenamefont {Pan}}]{q-netw-satellite_ycllzrclldllgxllyjljrhzzwczlclspwp17}%
  \BibitemOpen
  \bibfield  {author} {\bibinfo {author} {\bibfnamefont {J.}~\bibnamefont {Yin}}, \bibinfo {author} {\bibfnamefont {Y.}~\bibnamefont {Cao}}, \bibinfo {author} {\bibfnamefont {Y.-H.}\ \bibnamefont {Li}}, \bibinfo {author} {\bibfnamefont {S.-K.}\ \bibnamefont {Liao}}, \bibinfo {author} {\bibfnamefont {L.}~\bibnamefont {Zhang}}, \bibinfo {author} {\bibfnamefont {J.-G.}\ \bibnamefont {Ren}}, \bibinfo {author} {\bibfnamefont {W.-Q.}\ \bibnamefont {Cai}}, \bibinfo {author} {\bibfnamefont {W.-Y.}\ \bibnamefont {Liu}}, \bibinfo {author} {\bibfnamefont {B.}~\bibnamefont {Li}}, \bibinfo {author} {\bibfnamefont {H.}~\bibnamefont {Dai}}, \bibinfo {author} {\bibfnamefont {G.-B.}\ \bibnamefont {Li}}, \bibinfo {author} {\bibfnamefont {Q.-M.}\ \bibnamefont {Lu}}, \bibinfo {author} {\bibfnamefont {Y.-H.}\ \bibnamefont {Gong}}, \bibinfo {author} {\bibfnamefont {Y.}~\bibnamefont {Xu}}, \bibinfo {author} {\bibfnamefont {S.-L.}\ \bibnamefont {Li}}, \bibinfo {author} {\bibfnamefont {F.-Z.}\ \bibnamefont {Li}}, \bibinfo {author}
  {\bibfnamefont {Y.-Y.}\ \bibnamefont {Yin}}, \bibinfo {author} {\bibfnamefont {Z.-Q.}\ \bibnamefont {Jiang}}, \bibinfo {author} {\bibfnamefont {M.}~\bibnamefont {Li}}, \bibinfo {author} {\bibfnamefont {J.-J.}\ \bibnamefont {Jia}}, \bibinfo {author} {\bibfnamefont {G.}~\bibnamefont {Ren}}, \bibinfo {author} {\bibfnamefont {D.}~\bibnamefont {He}}, \bibinfo {author} {\bibfnamefont {Y.-L.}\ \bibnamefont {Zhou}}, \bibinfo {author} {\bibfnamefont {X.-X.}\ \bibnamefont {Zhang}}, \bibinfo {author} {\bibfnamefont {N.}~\bibnamefont {Wang}}, \bibinfo {author} {\bibfnamefont {X.}~\bibnamefont {Chang}}, \bibinfo {author} {\bibfnamefont {Z.-C.}\ \bibnamefont {Zhu}}, \bibinfo {author} {\bibfnamefont {N.-L.}\ \bibnamefont {Liu}}, \bibinfo {author} {\bibfnamefont {Y.-A.}\ \bibnamefont {Chen}}, \bibinfo {author} {\bibfnamefont {C.-Y.}\ \bibnamefont {Lu}}, \bibinfo {author} {\bibfnamefont {R.}~\bibnamefont {Shu}}, \bibinfo {author} {\bibfnamefont {C.-Z.}\ \bibnamefont {Peng}}, \bibinfo {author} {\bibfnamefont {J.-Y.}\
  \bibnamefont {Wang}},\ and\ \bibinfo {author} {\bibfnamefont {J.-W.}\ \bibnamefont {Pan}},\ }\bibfield  {title} {\bibinfo {title} {Satellite-based entanglement distribution over 1200 kilometers},\ }\href {https://doi.org/10.1126/science.aan3211} {\bibfield  {journal} {\bibinfo  {journal} {Science}\ }\textbf {\bibinfo {volume} {356}},\ \bibinfo {pages} {1140} (\bibinfo {year} {2017})}\BibitemShut {NoStop}%
\bibitem [{\citenamefont {Nokkala}\ \emph {et~al.}(2024)\citenamefont {Nokkala}, \citenamefont {Piilo},\ and\ \citenamefont {Bianconi}}]{q-netw_npb23}%
  \BibitemOpen
  \bibfield  {author} {\bibinfo {author} {\bibfnamefont {J.}~\bibnamefont {Nokkala}}, \bibinfo {author} {\bibfnamefont {J.}~\bibnamefont {Piilo}},\ and\ \bibinfo {author} {\bibfnamefont {G.}~\bibnamefont {Bianconi}},\ }\bibfield  {title} {\bibinfo {title} {Complex quantum networks: A topical review},\ }\href {https://doi.org/10.1088/1751-8121/ad41a6} {\bibfield  {journal} {\bibinfo  {journal} {Journal of Physics A: Mathematical and Theoretical}\ }\textbf {\bibinfo {volume} {57}},\ \bibinfo {pages} {233001} (\bibinfo {year} {2024})}\BibitemShut {NoStop}%
\bibitem [{\citenamefont {Chepuri}\ and\ \citenamefont {Kov{\'a}cs}(2023)}]{Chepuri2023}%
  \BibitemOpen
  \bibfield  {author} {\bibinfo {author} {\bibfnamefont {R.~T.~C.}\ \bibnamefont {Chepuri}}\ and\ \bibinfo {author} {\bibfnamefont {I.~A.}\ \bibnamefont {Kov{\'a}cs}},\ }\bibfield  {title} {\bibinfo {title} {Complex quantum network models from spin clusters},\ }\href {https://doi.org/10.1038/s42005-023-01394-8} {\bibfield  {journal} {\bibinfo  {journal} {Communications Physics}\ }\textbf {\bibinfo {volume} {6}},\ \bibinfo {pages} {271} (\bibinfo {year} {2023})}\BibitemShut {NoStop}%
\bibitem [{\citenamefont {Orieux}\ and\ \citenamefont {Diamanti}(2016)}]{Orieux2016}%
  \BibitemOpen
  \bibfield  {author} {\bibinfo {author} {\bibfnamefont {A.}~\bibnamefont {Orieux}}\ and\ \bibinfo {author} {\bibfnamefont {E.}~\bibnamefont {Diamanti}},\ }\bibfield  {title} {\bibinfo {title} {Recent advances on integrated quantum communications},\ }\href {https://doi.org/10.1088/2040-8978/18/8/083002} {\bibfield  {journal} {\bibinfo  {journal} {Journal of Optics}\ }\textbf {\bibinfo {volume} {18}},\ \bibinfo {pages} {083002} (\bibinfo {year} {2016})}\BibitemShut {NoStop}%
\bibitem [{\citenamefont {Razavi}(2018)}]{Razavi2018}%
  \BibitemOpen
  \bibfield  {author} {\bibinfo {author} {\bibfnamefont {M.}~\bibnamefont {Razavi}},\ }\href {https://doi.org/10.1088/978-1-6817-4653-1} {\emph {\bibinfo {title} {An Introduction to Quantum Communications Networks: Or, how shall we communicate in the quantum era?}}}\ (\bibinfo  {publisher} {Morgan \& Claypool Publishers},\ \bibinfo {year} {2018})\BibitemShut {NoStop}%
\bibitem [{\citenamefont {{de Forges de Parny}}\ \emph {et~al.}(2023)\citenamefont {{de Forges de Parny}}, \citenamefont {Alibart}, \citenamefont {Debaud}, \citenamefont {Gressani}, \citenamefont {Lagarrigue}, \citenamefont {Martin}, \citenamefont {Metrat}, \citenamefont {Schiavon}, \citenamefont {Troisi}, \citenamefont {Diamanti}, \citenamefont {G{\'e}lard}, \citenamefont {Kerstel}, \citenamefont {Tanzilli},\ and\ \citenamefont {Van Den~Bossche}}]{q-netw-satellite_fpadglmmstdgktvdb23}%
  \BibitemOpen
  \bibfield  {author} {\bibinfo {author} {\bibfnamefont {L.}~\bibnamefont {{de Forges de Parny}}}, \bibinfo {author} {\bibfnamefont {O.}~\bibnamefont {Alibart}}, \bibinfo {author} {\bibfnamefont {J.}~\bibnamefont {Debaud}}, \bibinfo {author} {\bibfnamefont {S.}~\bibnamefont {Gressani}}, \bibinfo {author} {\bibfnamefont {A.}~\bibnamefont {Lagarrigue}}, \bibinfo {author} {\bibfnamefont {A.}~\bibnamefont {Martin}}, \bibinfo {author} {\bibfnamefont {A.}~\bibnamefont {Metrat}}, \bibinfo {author} {\bibfnamefont {M.}~\bibnamefont {Schiavon}}, \bibinfo {author} {\bibfnamefont {T.}~\bibnamefont {Troisi}}, \bibinfo {author} {\bibfnamefont {E.}~\bibnamefont {Diamanti}}, \bibinfo {author} {\bibfnamefont {P.}~\bibnamefont {G{\'e}lard}}, \bibinfo {author} {\bibfnamefont {E.}~\bibnamefont {Kerstel}}, \bibinfo {author} {\bibfnamefont {S.}~\bibnamefont {Tanzilli}},\ and\ \bibinfo {author} {\bibfnamefont {M.}~\bibnamefont {Van Den~Bossche}},\ }\bibfield  {title} {\bibinfo {title} {Satellite-based quantum information networks:
  Use cases, architecture, and roadmap},\ }\href {https://doi.org/10.1038/s42005-022-01123-7} {\bibfield  {journal} {\bibinfo  {journal} {Communications Physics}\ }\textbf {\bibinfo {volume} {6}},\ \bibinfo {pages} {1} (\bibinfo {year} {2023})}\BibitemShut {NoStop}%
\bibitem [{\citenamefont {Christensen}(2002)}]{percolation-theor_c02}%
  \BibitemOpen
  \bibfield  {author} {\bibinfo {author} {\bibfnamefont {K.}~\bibnamefont {Christensen}},\ }\href@noop {} {\bibinfo {title} {Percolation {{Theory}}}} (\bibinfo {year} {2002})\BibitemShut {NoStop}%
\bibitem [{\citenamefont {Burda}\ \emph {et~al.}(2001)\citenamefont {Burda}, \citenamefont {Correia},\ and\ \citenamefont {Krzywicki}}]{random-tree_bck01}%
  \BibitemOpen
  \bibfield  {author} {\bibinfo {author} {\bibfnamefont {Z.}~\bibnamefont {Burda}}, \bibinfo {author} {\bibfnamefont {J.~D.}\ \bibnamefont {Correia}},\ and\ \bibinfo {author} {\bibfnamefont {A.}~\bibnamefont {Krzywicki}},\ }\bibfield  {title} {\bibinfo {title} {Statistical ensemble of scale-free random graphs},\ }\href {https://doi.org/10.1103/PhysRevE.64.046118} {\bibfield  {journal} {\bibinfo  {journal} {Physical Review E}\ }\textbf {\bibinfo {volume} {64}},\ \bibinfo {pages} {046118} (\bibinfo {year} {2001})}\BibitemShut {NoStop}%
\bibitem [{\citenamefont {Smoluchowski}(1916)}]{smoluchowski_s16}%
  \BibitemOpen
  \bibfield  {author} {\bibinfo {author} {\bibfnamefont {M.~V.}\ \bibnamefont {Smoluchowski}},\ }\bibfield  {title} {\bibinfo {title} {Drei {{Vortrage}} uber {{Diffusion}}, {{Brownsche Bewegung}} und {{Koagulation}} von {{Kolloidteilchen}}},\ }\href@noop {} {\bibfield  {journal} {\bibinfo  {journal} {Zeitschrift fur Physik}\ }\textbf {\bibinfo {volume} {17}},\ \bibinfo {pages} {557} (\bibinfo {year} {1916})}\BibitemShut {NoStop}%
\bibitem [{\citenamefont {Bonato}\ \emph {et~al.}(2009)\citenamefont {Bonato}, \citenamefont {Tomaello}, \citenamefont {Deppo}, \citenamefont {Naletto},\ and\ \citenamefont {Villoresi}}]{qkd-satellite_btdnv09}%
  \BibitemOpen
  \bibfield  {author} {\bibinfo {author} {\bibfnamefont {C.}~\bibnamefont {Bonato}}, \bibinfo {author} {\bibfnamefont {A.}~\bibnamefont {Tomaello}}, \bibinfo {author} {\bibfnamefont {V.~D.}\ \bibnamefont {Deppo}}, \bibinfo {author} {\bibfnamefont {G.}~\bibnamefont {Naletto}},\ and\ \bibinfo {author} {\bibfnamefont {P.}~\bibnamefont {Villoresi}},\ }\bibfield  {title} {\bibinfo {title} {Feasibility of satellite quantum key distribution},\ }\href {https://doi.org/10.1088/1367-2630/11/4/045017} {\bibfield  {journal} {\bibinfo  {journal} {New Journal of Physics}\ }\textbf {\bibinfo {volume} {11}},\ \bibinfo {pages} {045017} (\bibinfo {year} {2009})}\BibitemShut {NoStop}%
\bibitem [{\citenamefont {Cao}\ \emph {et~al.}(2013)\citenamefont {Cao}, \citenamefont {Hong}, \citenamefont {Du},\ and\ \citenamefont {Zhang}}]{link-add_chdz13}%
  \BibitemOpen
  \bibfield  {author} {\bibinfo {author} {\bibfnamefont {X.-B.}\ \bibnamefont {Cao}}, \bibinfo {author} {\bibfnamefont {C.}~\bibnamefont {Hong}}, \bibinfo {author} {\bibfnamefont {W.-B.}\ \bibnamefont {Du}},\ and\ \bibinfo {author} {\bibfnamefont {J.}~\bibnamefont {Zhang}},\ }\bibfield  {title} {\bibinfo {title} {Improving the network robustness against cascading failures by adding links},\ }\href {https://doi.org/10.1016/j.chaos.2013.08.007} {\bibfield  {journal} {\bibinfo  {journal} {Chaos, Solitons \& Fractals}\ }\textbf {\bibinfo {volume} {57}},\ \bibinfo {pages} {35} (\bibinfo {year} {2013})}\BibitemShut {NoStop}%
\bibitem [{\citenamefont {Chen}\ \emph {et~al.}(2022)\citenamefont {Chen}, \citenamefont {Zhao}, \citenamefont {Qin}, \citenamefont {Meng},\ and\ \citenamefont {Gao}}]{link-add_czqmg22}%
  \BibitemOpen
  \bibfield  {author} {\bibinfo {author} {\bibfnamefont {C.-Y.}\ \bibnamefont {Chen}}, \bibinfo {author} {\bibfnamefont {Y.}~\bibnamefont {Zhao}}, \bibinfo {author} {\bibfnamefont {H.}~\bibnamefont {Qin}}, \bibinfo {author} {\bibfnamefont {X.}~\bibnamefont {Meng}},\ and\ \bibinfo {author} {\bibfnamefont {J.}~\bibnamefont {Gao}},\ }\bibfield  {title} {\bibinfo {title} {Robustness of interdependent scale-free networks based on link addition strategies},\ }\href {https://doi.org/10.1016/j.physa.2022.127851} {\bibfield  {journal} {\bibinfo  {journal} {Physica A: Statistical Mechanics and its Applications}\ }\textbf {\bibinfo {volume} {604}},\ \bibinfo {pages} {127851} (\bibinfo {year} {2022})}\BibitemShut {NoStop}%
\bibitem [{\citenamefont {Wu}\ \emph {et~al.}(2022)\citenamefont {Wu}, \citenamefont {Meng}, \citenamefont {Danziger}, \citenamefont {Cornelius}, \citenamefont {Tian},\ and\ \citenamefont {Barab{\'a}si}}]{interdepend-recover_wmdctb22}%
  \BibitemOpen
  \bibfield  {author} {\bibinfo {author} {\bibfnamefont {H.}~\bibnamefont {Wu}}, \bibinfo {author} {\bibfnamefont {X.}~\bibnamefont {Meng}}, \bibinfo {author} {\bibfnamefont {M.~M.}\ \bibnamefont {Danziger}}, \bibinfo {author} {\bibfnamefont {S.~P.}\ \bibnamefont {Cornelius}}, \bibinfo {author} {\bibfnamefont {H.}~\bibnamefont {Tian}},\ and\ \bibinfo {author} {\bibfnamefont {A.-L.}\ \bibnamefont {Barab{\'a}si}},\ }\bibfield  {title} {\bibinfo {title} {Fragmentation of outage clusters during the recovery of power distribution grids},\ }\href {https://doi.org/10.1038/s41467-022-35104-9} {\bibfield  {journal} {\bibinfo  {journal} {Nature Communications}\ }\textbf {\bibinfo {volume} {13}},\ \bibinfo {pages} {7372} (\bibinfo {year} {2022})}\BibitemShut {NoStop}%
\bibitem [{\citenamefont {Ma}\ \emph {et~al.}(2023)\citenamefont {Ma}, \citenamefont {Shen}, \citenamefont {Ma}, \citenamefont {Hu},\ and\ \citenamefont {Peng}}]{link-add_msmhp23}%
  \BibitemOpen
  \bibfield  {author} {\bibinfo {author} {\bibfnamefont {S.}~\bibnamefont {Ma}}, \bibinfo {author} {\bibfnamefont {B.}~\bibnamefont {Shen}}, \bibinfo {author} {\bibfnamefont {J.}~\bibnamefont {Ma}}, \bibinfo {author} {\bibfnamefont {W.}~\bibnamefont {Hu}},\ and\ \bibinfo {author} {\bibfnamefont {T.}~\bibnamefont {Peng}},\ }\bibfield  {title} {\bibinfo {title} {Improvement of network robustness against cascading failures based on the min\textendash max edge-adding strategy},\ }\href {https://doi.org/10.1016/j.physa.2022.128442} {\bibfield  {journal} {\bibinfo  {journal} {Physica A: Statistical Mechanics and its Applications}\ }\textbf {\bibinfo {volume} {611}},\ \bibinfo {pages} {128442} (\bibinfo {year} {2023})}\BibitemShut {NoStop}%
\bibitem [{\citenamefont {Hu}\ \emph {et~al.}(2024)\citenamefont {Hu}, \citenamefont {Dong}, \citenamefont {Lambiotte}, \citenamefont {Christensen}, \citenamefont {Fan}, \citenamefont {Tian}, \citenamefont {Gao}, \citenamefont {Havlin},\ and\ \citenamefont {Meng}}]{q-netw-long-path_hdlcftghm24}%
  \BibitemOpen
  \bibfield  {author} {\bibinfo {author} {\bibfnamefont {X.}~\bibnamefont {Hu}}, \bibinfo {author} {\bibfnamefont {G.}~\bibnamefont {Dong}}, \bibinfo {author} {\bibfnamefont {R.}~\bibnamefont {Lambiotte}}, \bibinfo {author} {\bibfnamefont {K.}~\bibnamefont {Christensen}}, \bibinfo {author} {\bibfnamefont {J.}~\bibnamefont {Fan}}, \bibinfo {author} {\bibfnamefont {L.}~\bibnamefont {Tian}}, \bibinfo {author} {\bibfnamefont {J.}~\bibnamefont {Gao}}, \bibinfo {author} {\bibfnamefont {S.}~\bibnamefont {Havlin}},\ and\ \bibinfo {author} {\bibfnamefont {X.}~\bibnamefont {Meng}},\ }\href {https://doi.org/10.48550/arXiv.2402.15462} {\bibinfo {title} {Unveiling the {{Importance}} of {{Longer Paths}} in {{Quantum Networks}}}} (\bibinfo {year} {2024}),\ \Eprint {https://arxiv.org/abs/2402.15462} {arxiv:2402.15462} \BibitemShut {NoStop}%
\bibitem [{\citenamefont {Kim}\ and\ \citenamefont {Radicchi}(2024)}]{kim2024}%
  \BibitemOpen
  \bibfield  {author} {\bibinfo {author} {\bibfnamefont {M.}~\bibnamefont {Kim}}\ and\ \bibinfo {author} {\bibfnamefont {F.}~\bibnamefont {Radicchi}},\ }\href@noop {} {\bibinfo {title} {Shortest-path percolation on complex networks}} (\bibinfo {year} {2024}),\ \Eprint {https://arxiv.org/abs/2402.06753} {arXiv:2402.06753 [physics.soc-ph]} \BibitemShut {NoStop}%
\bibitem [{\citenamefont {Wilson}(1996)}]{wilsonGeneratingRandomSpanning1996}%
  \BibitemOpen
  \bibfield  {author} {\bibinfo {author} {\bibfnamefont {D.~B.}\ \bibnamefont {Wilson}},\ }\bibfield  {title} {\bibinfo {title} {Generating random spanning trees more quickly than the cover time},\ }in\ \href@noop {} {\emph {\bibinfo {booktitle} {Proceedings of the Twenty-Eighth Annual {{ACM}} Symposium on {{Theory}} of Computing - {{STOC}} '96}}}\ (\bibinfo  {publisher} {ACM Press},\ \bibinfo {year} {1996})\ pp.\ \bibinfo {pages} {296--303}\BibitemShut {NoStop}%
\bibitem [{\citenamefont {Aldous}(1990)}]{aldousRandomWalkConstruction1990}%
  \BibitemOpen
  \bibfield  {author} {\bibinfo {author} {\bibfnamefont {D.~J.}\ \bibnamefont {Aldous}},\ }\bibfield  {title} {\bibinfo {title} {The {{Random Walk Construction}} of {{Uniform Spanning Trees}} and {{Uniform Labelled Trees}}},\ }\href@noop {} {\bibfield  {journal} {\bibinfo  {journal} {SIAM Journal on Discrete Mathematics}\ }\textbf {\bibinfo {volume} {3}},\ \bibinfo {pages} {450} (\bibinfo {year} {1990})}\BibitemShut {NoStop}%
\bibitem [{\citenamefont {Aldous}(1991{\natexlab{a}})}]{random-tree-contin_a91}%
  \BibitemOpen
  \bibfield  {author} {\bibinfo {author} {\bibfnamefont {D.}~\bibnamefont {Aldous}},\ }\bibfield  {title} {\bibinfo {title} {The {{Continuum Random Tree}}. {{I}}},\ }\href {https://doi.org/10.1214/aop/1176990534} {\bibfield  {journal} {\bibinfo  {journal} {The Annals of Probability}\ }\textbf {\bibinfo {volume} {19}},\ \bibinfo {pages} {1} (\bibinfo {year} {1991}{\natexlab{a}})}\BibitemShut {NoStop}%
\bibitem [{\citenamefont {Aldous}(1991{\natexlab{b}})}]{random-tree-contin_a91a}%
  \BibitemOpen
  \bibfield  {author} {\bibinfo {author} {\bibfnamefont {D.}~\bibnamefont {Aldous}},\ }\bibfield  {title} {\bibinfo {title} {The {{Continuum}} random tree {{II}}: An overview},\ }in\ \href@noop {} {\emph {\bibinfo {booktitle} {Stochastic {{Analysis}}: {{Proceedings}} of the {{Durham Symposium}} on {{Stochastic Analysis}}, 1990}}},\ \bibinfo {series and number} {London {{Mathematical Society Lecture Note Series}}},\ \bibinfo {editor} {edited by\ \bibinfo {editor} {\bibfnamefont {M.~T.}\ \bibnamefont {Barlow}}\ and\ \bibinfo {editor} {\bibfnamefont {N.~H.}\ \bibnamefont {Bingham}}}\ (\bibinfo  {publisher} {{Cambridge University Press}},\ \bibinfo {address} {{Cambridge}},\ \bibinfo {year} {1991})\ pp.\ \bibinfo {pages} {23--70}\BibitemShut {NoStop}%
\bibitem [{\citenamefont {Aldous}(1993)}]{random-tree-contin_a93}%
  \BibitemOpen
  \bibfield  {author} {\bibinfo {author} {\bibfnamefont {D.}~\bibnamefont {Aldous}},\ }\bibfield  {title} {\bibinfo {title} {The {{Continuum Random Tree III}}},\ }\href {https://doi.org/10.1214/aop/1176989404} {\bibfield  {journal} {\bibinfo  {journal} {The Annals of Probability}\ }\textbf {\bibinfo {volume} {21}},\ \bibinfo {pages} {248} (\bibinfo {year} {1993})}\BibitemShut {NoStop}%
\bibitem [{\citenamefont {Peres}\ and\ \citenamefont {Revelle}(2005)}]{random-tree-rayleigh_pr05}%
  \BibitemOpen
  \bibfield  {author} {\bibinfo {author} {\bibfnamefont {Y.}~\bibnamefont {Peres}}\ and\ \bibinfo {author} {\bibfnamefont {D.}~\bibnamefont {Revelle}},\ }\href {https://doi.org/10.48550/arXiv.math/0410430} {\bibinfo {title} {Scaling limits of the uniform spanning tree and loop-erased random walk on finite graphs}} (\bibinfo {year} {2005}),\ \Eprint {https://arxiv.org/abs/math/0410430} {arxiv:math/0410430} \BibitemShut {NoStop}%
\bibitem [{\citenamefont {Kolchin}(1977)}]{random-tree_k77}%
  \BibitemOpen
  \bibfield  {author} {\bibinfo {author} {\bibfnamefont {V.~F.}\ \bibnamefont {Kolchin}},\ }\bibfield  {title} {\bibinfo {title} {Branching processes, random trees, and a generalized scheme of arrangements of particles},\ }\href {https://doi.org/10.1007/BF01788236} {\bibfield  {journal} {\bibinfo  {journal} {Mathematical notes of the Academy of Sciences of the USSR}\ }\textbf {\bibinfo {volume} {21}},\ \bibinfo {pages} {386} (\bibinfo {year} {1977})}\BibitemShut {NoStop}%
\bibitem [{\citenamefont {Hladk{\'y}}\ \emph {et~al.}(2018)\citenamefont {Hladk{\'y}}, \citenamefont {Nachmias},\ and\ \citenamefont {Tran}}]{random-tree_hnt18}%
  \BibitemOpen
  \bibfield  {author} {\bibinfo {author} {\bibfnamefont {J.}~\bibnamefont {Hladk{\'y}}}, \bibinfo {author} {\bibfnamefont {A.}~\bibnamefont {Nachmias}},\ and\ \bibinfo {author} {\bibfnamefont {T.}~\bibnamefont {Tran}},\ }\bibfield  {title} {\bibinfo {title} {The {{Local Limit}} of the {{Uniform Spanning Tree}} on {{Dense Graphs}}},\ }\href {https://doi.org/10.1007/s10955-017-1933-5} {\bibfield  {journal} {\bibinfo  {journal} {Journal of Statistical Physics}\ }\textbf {\bibinfo {volume} {173}},\ \bibinfo {pages} {502} (\bibinfo {year} {2018})}\BibitemShut {NoStop}%
\bibitem [{\citenamefont {Cressie}\ and\ \citenamefont {Borkent}(1986)}]{fract-deriv_cb86}%
  \BibitemOpen
  \bibfield  {author} {\bibinfo {author} {\bibfnamefont {N.}~\bibnamefont {Cressie}}\ and\ \bibinfo {author} {\bibfnamefont {M.}~\bibnamefont {Borkent}},\ }\bibfield  {title} {\bibinfo {title} {The moment generating function has its moments},\ }\href {https://doi.org/10.1016/0378-3758(86)90143-6} {\bibfield  {journal} {\bibinfo  {journal} {Journal of Statistical Planning and Inference}\ }\textbf {\bibinfo {volume} {13}},\ \bibinfo {pages} {337} (\bibinfo {year} {1986})}\BibitemShut {NoStop}%
\end{thebibliography}%

\clearpage
\newpage
\appendix

\renewcommand{\appendixname}{Section}
\renewcommand{\thesection}{\arabic{section}}

\section{Random tree characteristics}

\begin{figure}[t!]
    \centering
    \begin{minipage}[b]{110pt}
		\centering
        {\subcaption{\label{fig_tree_illustration}}\includegraphics[width=110pt]{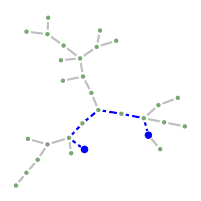}}
    \end{minipage} 
    \begin{minipage}[b]{110pt}
		\centering
        {\subcaption{\label{fig_tree_single_remove}}\includegraphics[width=110pt]{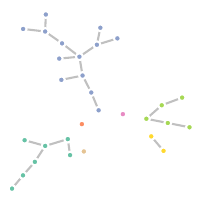}}
    \end{minipage} 
    \begin{minipage}[b]{240pt}
		\centering
        {\subcaption{\label{fig_rayleigh}}\includegraphics[width=240pt]{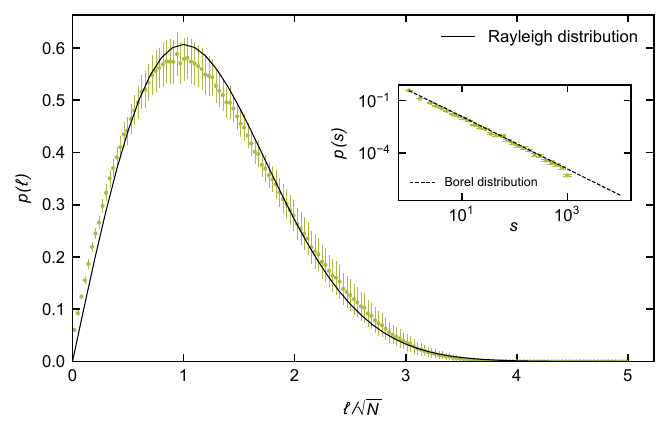}}
    \end{minipage} 
    \vspace{-3mm}
    \caption{ 
    Single path removal on the uniform spanning tree (UST). \subref{fig_tree_illustration}~We randomly and uniformly pick two users (blue) connected by a path (dashed blue) from a UST of 30 users.
    \subref{fig_tree_single_remove}~The path connecting the two users is removed, fragmenting the UST into several components shown by different colors.
    \subref{fig_rayleigh}~The path length distribution in the UST follows the Rayleigh distribution ($N=1000$). The inset illustrates the component size distribution after the removal of a single path, adhering to the Borel distribution.
    \hfill\hfill}
\end{figure}

Here, we consider uniform spanning trees (UST) constructed on $N$ nodes using Wilson's algorithm \cite{wilsonGeneratingRandomSpanning1996}, but we have also confirmed that the results remain the same using the Aldous--Boulder algorithm \cite{aldousRandomWalkConstruction1990}.
For any tree [Fig.~\ref{fig_tree_illustration}], the removal of any path of length $l$ leads to the fragmentation of the network into $l+1$ disconnected components [Fig.~\ref{fig_tree_single_remove}].
Given two randomly chosen nodes in a UST of size $N$, the path length $l$ between them follows the Rayleigh distribution~\cite{random-tree-contin_a91,*random-tree-contin_a91a,*random-tree-contin_a93,*random-tree-rayleigh_pr05} in the large $N$ (continuum) limit,
\begin{equation}
    p_\text{UST}(l|N)\simeq \left(l/\sigma^2\right)e^{-l^2/\left(2\sigma^2\right)},
\end{equation}
where $\sigma\sim \sqrt{N}$. For USTs sampled on a complete graph $K_{N}$, one expects exactly $\sigma=\sqrt{N}$, corresponding to the Brownian continuum random tree~\cite{random-tree-contin_a91,*random-tree-contin_a91a,*random-tree-contin_a93,*random-tree-rayleigh_pr05}, which agrees with our simulations [Fig.~\ref{fig_rayleigh}].
Thus, the average number of components resulting from removing a path is
\begin{equation}
\label{eq_rayleigh_mean}
    \int_0^{\infty} dl \left(l+1\right) p_\text{UST}(l|N) = \sqrt{\frac{\pi }{2}} \sigma +1 \sim  \sqrt{N}.
\end{equation}

Furthermore, any fragmented components of a UST can still be regarded as random trees, resulting from a Galton--Watson branching process~\cite{random-tree_k77,*random-tree_hnt18}. Then, the component size distribution follows critical Borel distribution (in the $N \to \infty$ limit), yielding 
\begin{equation}
\label{eq_borel}
p_\text{UST}(s) \sim s^{-3/2}
\end{equation}
for large $s$.
As for comparing $p_\text{UST}(s)$ with simulations,
we note that the $3/2$ exponent on $s$ is small enough to disable self-averaging. Hence, we cannot test Eq.~\eqref{eq_borel} directly. Instead, we employ a different approach: we count the number of components for each size $s$ after fragmenting the original UST, and then divide each component count for size $s$ by a global factor $l+1$, which is the total number of components resulting from the fragmentation.
This is to suppress the fluctuation in $p_\text{UST}(s)$ resulting from very small components, which can happen when the number of fragmented components $l+1$ is large.
Our simulation analysis also agrees with the Borel distribution for USTs [inset of Fig.~\ref{fig_rayleigh}].

Finally, we consider $r(s|s')$, the average number of $s$-components yielded by fragmenting an $s'$-component, assuming that the $s'$-component is a UST. Note that a UST of size $s'$ typically results in $\sqrt{s'}$ components [Eq.~\eqref{eq_rayleigh_mean}]. Therefore, we have
\begin{equation}
\label{eq_borel_si}
r(s|s')\sim {\left(s'\right)^{1/2}}{s^{-3/2}}.
\end{equation}

\section{Generalized Smoluchowski equation}

Let $v(s)$ be the number of $s$-components per site, namely, $N v(s)$ is the total number of components of size $s$ in the network. The rate equation for $v(s)$ is given by:
\begin{eqnarray}
\label{eq_markov}
    &&N v(s,t+1)\nonumber\\
    &=&N v(s,t) - \left[\left\langle s\right\rangle-1\right]^{-1} s\left(s-1\right)v(s,t)\nonumber\\
    &+& \left[\left\langle s\right\rangle-1\right]^{-1} \sum_{s'=s+1}^{\infty} s'\left(s'-1\right)  v(s',t)r(s|s')\nonumber\\
    &+&\alpha \sum_{s'=1}^{s-1} 
    s'  v(s',t) \left(s-s'\right) \frac{N v(s-s',t)-\delta(s-2s')}{N-s'} \nonumber\\
    &-&\alpha 
    \sum_{s'=1}^{\infty}s' v(s',t) \left( \delta(s-s')  +  s \frac{N v(s,t)-\delta(s-s')}{N-s'}\right),\nonumber\\
\end{eqnarray}
where $\left\langle s\right\rangle = \sum_{s=1}^{\infty} s^2  v(s,t)$ is the mean component size per site.
The second term on the RHS of Eq.~\eqref{eq_markov_approx} is the probability of selecting an $s$-component and removing a path from it. The third term is the rate of obtaining an $s$-component by removing a path from a larger $s'$-component, for which the conditional rate $r(s|s')$ is the average number of $s$-components yielded from the $s'$-component. The fourth term is the rate of picking two components of sizes $s'$ and $s-s'$ and merging them into a new $s$-component, and the last term is the rate of picking an $s$-component for merging.

When $\alpha$ is small, we expect that $v(\infty, t)\to 0$ as $t\to \infty$. Thus, for $1 \ll s\ll N$,
Eq.~\eqref{eq_markov} can be approximated as 
\begin{eqnarray}
\label{eq_markov_approx}
    N v(s,t+1)&=&N v(s,t) - \left\langle s\right\rangle^{-1} 
    s^2 
    v(s,t)\nonumber\\
    &+&\left\langle s\right\rangle^{-1} \sum_{s'=s+1}^{\infty} 
    \left(s'\right)^2 v(s',t)r(s|s')\nonumber\\
    &+&\alpha \sum_{s'=1}^{s-1} s'  v(s',t) \left(s-s'\right) v(s-s',t) \nonumber\\
    &-&2 \alpha  s v(s,t),
\end{eqnarray}
where
\begin{equation}
    r(s|s') \simeq k {\left(s'\right)^{1/2}}{s^{-3/2}},
\end{equation}
as given by Eq.~\eqref{eq_borel_si} in Section~1.
The normalization coefficient $k$ is to be fixed by the continuity condition of the Smoluchowski equation.

To find the steady state, we introduce
\begin{equation}
    f(s)=\begin{cases} 
      s v(s,\infty) & s>0 \\
      0 & s\le 0
\end{cases},
\end{equation}
which is the probability of selecting a node that belongs to an $s$-component in the steady state. From Eq.~\eqref{eq_markov_approx}, we have
\begin{eqnarray}
\label{eq_markov_steady}
    &&sf(s)-
    k s^{-3/2} \sum_{s'=s+1}^{\infty} \left(s' \right)^{3/2}f(s')\nonumber\\
    &=& \alpha \left\langle s\right\rangle \left(\sum_{s'=1}^{s-1} f(s')f(s-s') - 2f(s) \right)
\end{eqnarray}
for $s>0$. Next, we introduce an auxiliary function $g(s)$:
\begin{equation}
    g(s)=\begin{cases} 
      s^{-3/2} \sum_{s'=s}^{\infty} \left(s' \right)^{3/2}f(s') & s>0 \\
      0 & s\le 0
\end{cases},
\end{equation}
as well as the generating functions (Laplace transforms) of both $f(s)$ and $g(s)$: 
\begin{eqnarray}
X(t)&=&\sum_{s=0}^{\infty} f(s) e^{s t}=\sum_{s=-\infty}^{\infty} f(s) z^{-s},\nonumber\\
Z(t)&=&\sum_{s=0}^{\infty} g(s) e^{s t}=\sum_{s=-\infty}^{\infty} g(s) z^{-s},
\end{eqnarray}
where $z=e^{-t}$ corresponds to the traditional z-variable in the Z transform.
The Laplace transform of Eq.~\eqref{eq_markov_steady} reads as
\begin{eqnarray}
\label{eq_markov_steady_z}
X'(t) + k X(t) -k Z(t)= \alpha X'(0) \left(X^2(t)-2X(t) \right),\nonumber\\
\end{eqnarray}
where $Z(t)$ follows the fractional differential equation, 
\begin{eqnarray}
\label{eq_markov_steady_z_fract}
\left(1-e^{-t}\right) D^{\left(3/2\right)}_{t} Z(t)=  D^{\left(3/2\right)}_{t} X(t)- D^{\left(3/2\right)}_{t} X(0) .\nonumber\\
\end{eqnarray}
Here, $D^{(r)}_{t}$ is the Cressie--Borkent $r$-fractional derivative operator~\cite{fract-deriv_cb86} with respect to $t$, corresponding to
\begin{eqnarray}
D^{\left(3/2\right)}_{t} X(t)&=&\sum_{s=0}^{\infty} s^{3/2} f(s) e^{st},\nonumber\\
D^{\left(3/2\right)}_{t} Z(t)&=&\sum_{s=0}^{\infty} s^{3/2} g(s) e^{st}.
\end{eqnarray}

\section{Solving the Smoluchowski equation by moments}

The initial conditions of $X(t)$ are given by
\begin{eqnarray}
    X(0)&=&1,\nonumber\\
    X'(0)&=&\left\langle s\right\rangle,\nonumber\\
    X''(0)&=&\left\langle s^2\right\rangle,\nonumber\\
    X'''(0)&=&\left\langle s^3\right\rangle.\nonumber\\
    &\vdots&
\end{eqnarray}
With these, $k$ and $\alpha$ could be obtained from the series expansion of Eq.~\eqref{eq_markov_steady_z}:
\begin{eqnarray}
\label{eq_markov_steady_z_series}
X'(0) + k X(0) -k Z(0)&=& -\alpha X'(0),\nonumber\\
X''(0) + k X'(0) -k Z'(0)&=& 0,\nonumber\\
X'''(0) + k X''(0) -k Z''(0)&=& 2\alpha \left[X'(0)\right]^3,\nonumber\\
&\vdots&
\end{eqnarray}
One can also show that
\begin{eqnarray}
    Z(0)&=&\sum_{s=1}^{\infty }s^{-3/2} \sum_{s'=s}^{\infty} \left(s' \right)^{3/2}f(s')\nonumber\\
    &=&\sum_{s'=1}^{\infty }\left(\sum_{s=1}^{s'} s^{-3/2} \right)\left(s' \right)^{3/2}f(s')\nonumber\\
    &=&\sum_{s'=1}^{\infty } H_{s'}^{(3/2)} \left(s' \right)^{3/2}f(s')\nonumber\\
    &\simeq&\sum_{s'=1}^{\infty } \left(\zeta\left(\frac{3}{2}\right)\left(s' \right)^{3/2}-2s'+\frac{1}{2}+\cdots\right) f(s')\nonumber\\
    &\simeq& \zeta \left(\frac{3}{2}\right) D^{\left(3/2\right)}_{t} X(0) - 2 X'(0)\cdots
\end{eqnarray}
depends on the $(3/2)$-fractional moment of $X(t)$, which is coupled to all regular, increasingly large moments $X(0)$, $X'(0)$, $\cdots$, thus preventing solving Eq.~\eqref{eq_markov_steady_z} by truncating higher moments. 
Here, $H_{s}^{(r)}$ denotes the harmonic number, and $\zeta(s)$ is the Riemann zeta function.
Similarly, we obtain that
\begin{eqnarray}
    Z'(0)&=&\sum_{s'=1}^{\infty } H_{s'}^{(1/2)} \left(s' \right)^{3/2}f(s')\nonumber\\
   &\simeq& 2 X''(0)+ \zeta \left(\frac{1}{2}\right) D^{\left(3/2\right)}_{t} X(0)+\cdots.\nonumber\\
   Z''(0)&=&\sum_{s'=1}^{\infty } H_{s'}^{(-1/2)} \left(s' \right)^{3/2}f(s')\nonumber\\
   &\simeq& \frac{2}{3} X'''(0)+ \frac{1}{2} X''(0)+\cdots.\nonumber\\
\end{eqnarray}

Near the phase transition threshold $\alpha\to \alpha^*$, we expect a power-law component size distribution [as suggested in Fig.~\ref{fig_s_tree_large}], $v(s, \infty)\sim s^{-\tau}$, up to some cutoff $s_{\max}$. 
This power-law ansatz gives rise to
\begin{equation}
\label{eq_markov_ansatz}
    f(s)= \begin{cases} 
      s^{-\tau+1}/H_{s_{\max}}^{(\tau-1)} & 0<s\leq s_{\max} \\
      0 & \text{otherwise}
\end{cases},
\end{equation}
which allows us to write down the $(3/2)$-fractional moment of $X(z)$,
\begin{equation}
    D^{\left(3/2\right)}_{t} X(0)= {H_{s_{\max}}^{\left(\tau -\frac{5}{2}\right)}}/{H_{s_{\max}}^{(\tau -1)}},
\end{equation}
as well as 
\begin{eqnarray}
  X'(0)&=&{H_{s_{\max}}^{\left(\tau -2\right)}}/{H_{s_{\max}}^{(\tau -1)}},\nonumber\\
  X''(0)&=&{H_{s_{\max}}^{\left(\tau -3\right)}}/{H_{s_{\max}}^{(\tau -1)}}.
\end{eqnarray}

Together with Eq.~\eqref{eq_markov_steady_z_series}, we solve for $k$ and $\alpha^*$ and find:
\begin{eqnarray}
  k&=&\frac{X''(0)}{Z'(0)-X'(0)}\nonumber\\
  &=&\frac{H_{s_{\max}}^{(\tau -3)}}{2 H_{s_{\max}}^{(\tau -3)}+\zeta \left(\frac{1}{2}\right) H_{s_{\max}}^{\left(\tau -\frac{5}{2}\right)}-H_{s_{\max}}^{(\tau -2)}},\nonumber\\
  \alpha^*&=& k \frac{Z(0)-X(0)}{X'(0)}-1\nonumber\\
  &=&k\frac{\zeta \left(\frac{3}{2}\right) H_{s_{\max}}^{\left(\tau -\frac{5}{2}\right)}-2 H_{s_{\max}}^{(\tau -2)}-H_{s_{\max}}^{(\tau -1)}}{2 H_{s_{\max}}^{(\tau -2)}}-1,
\end{eqnarray}
which yields
\begin{eqnarray}
  k&\simeq &\frac{1}{2}-\frac{1}{2} \frac{4-\tau}{7-2 \tau } \zeta \left(\frac{1}{2}\right)\frac{1}{\sqrt{s_{\max}} }+\cdots,\nonumber\\
  \alpha^*&\simeq &\frac{3-\tau }{7-2 \tau }  \zeta \left(\frac{3}{2}\right) \sqrt{s_{\max}} \nonumber\\
  &&+\left[-2 -\frac{(4-\tau ) (3-\tau )}{(7-2 \tau )^2}\zeta \left(\frac{1}{2}\right) \zeta \left(\frac{3}{2}\right) \right]+\cdots.\nonumber\\
\end{eqnarray}
Therefore, $\alpha^*\sim \sqrt{s_{\max}}$, as long as
$\tau <3$. For $\tau \ge 3$, logarithmic corrections would appear in the dominant term of $\alpha$ and smear the square-root dependence on $s_{\max}$.

Finally, to fix $\tau$, we consider the third moment of $f(s)$ [fourth moment of $v(s)$] in Eq.~\eqref{eq_markov_steady_z_series}. On the LHS, we have
\begin{equation}
    X'''(0) + k X''(0) -k Z''(0) \sim s_{\max}^{5-\tau},
\end{equation}
and on the RHS,
\begin{equation}
    2\alpha \left[X'(0)\right]^3 \sim 
    \begin{cases} 
      \left(s_{\max}^{3-\tau} \right)^3 & \alpha\sim s_{\max}^0  \\
      s_{\max}^{1/2} \left(s_{\max}^{3-\tau} \right)^3 & \alpha\sim s_{\max}^{1/2}
   \end{cases}.
\end{equation}
Matching them, we find $\tau=2$ for small $\alpha$, but $\tau=9/4$ when $\alpha$ approaches the critical threshold $\alpha^*\sim \sqrt{s_{\max}}$.

Finally, assuming $s_{\max} \simeq N$, we arrive at the theoretical results in the main paper.

\section{Path length balance equation}

Alternatively, we can write down the path length distribution averaged over all components:
\begin{equation}
p(l) =\frac{\sum_s  s\left(s-1\right) v(s) p_\text{UST}(l|s)}{\sum_s  s\left(s-1\right) v(s)}.
\end{equation}
The factor $s\left(s-1\right)$ represents the probability weight of uniformly selecting two nodes at random and finding that they belong to the same component of size $s$.
Therefore, the average path length of an $s$-component, averaged over all components, is given by
\begin{eqnarray}
\overline{l}&=\int_0^{\infty}  l  p(l)  dl
&=\frac{\sum_s  s\left(s-1\right) v(s) \sqrt{{\pi}/{2}}\sqrt{s}}{\sum_s  s\left(s-1\right) v(s)}
\end{eqnarray}
according to Eq.~\eqref{eq_rayleigh_mean}.

Note that at the steady state, one must have $\overline{l} = \alpha$, as long as more then $\alpha+1$ components exist. Then, with the power-law ansatz $v(s)\sim s^{-\tau}$, 
we have
\begin{equation}
     \sqrt{\frac{\pi}{2}} \sum_{s=1}^{s_{\max}}s^{\frac{3}{2}-\tau}(s-1)  \approx \alpha^*  \sum_{s=1}^{s_{\max}}s^{1-\tau}(s-1),
\end{equation}
with a maximum component size given by $s_{\max}$. Approximating the sum with an integral and $s-1$ by $s$ leads to 
\begin{equation}
    \alpha^* \approx \frac{3-\tau}{7-2\tau}\sqrt{2\pi}\sqrt{s_{\max}},
\end{equation}
when $\tau<3$, following the same scaling as derived by the Smoluchowski equation. Note that the prefactor in $\alpha^*$ is again not expected to be precise, as fluctuations cannot be neglected close to the phase transition.

\end{document}